\documentclass[aip, amsmath,amssymb,reprint,twocolumn]{revtex4-1}
\pdfoutput=1

\usepackage{graphicx}
\usepackage{xcolor}
\usepackage{soul}

\usepackage{bm}
\usepackage{color}
\usepackage{hyperref}
\usepackage[separate-uncertainty=true]{siunitx}
\usepackage{verbatim}
\usepackage{placeins} 

\definecolor{bondiblue}{rgb}{0.0, 0.58, 0.71}
\definecolor{purple}{rgb}{0.5, 0., 0.5}
\hypersetup{pdfauthor={Ellis et al.},pdftitle={Machine learning using magnetic stochastic synapses}}

\newif\ifchangemarks 

\changemarkstrue 
\changemarksfalse 

\ifchangemarks

\newcommand{\new} [1]{\textcolor{cyan}{#1}}
\newcommand{\draftinfo} [1]{#1}

\else

\newcommand{\new}[1]{#1}
\newcommand{\draftinfo} [1]{\relax}

\fi

\newif\iflowqaulitygraphics 

\lowqaulitygraphicsfalse 

\iflowqaulitygraphics
  \usepackage{epstopdf}
  \DeclareGraphicsExtensions{.png,.pdf} 
\fi

\begin{document}

\title[\textbf{Magnetic Stochastic Synapses}]{Machine learning using magnetic stochastic synapses}

\author{Matthew O. A. Ellis}%
\thanks{These authors contributed equally to this work.}
\affiliation{Department of Computer Science, University of Sheffield, Sheffield, S1 4DP, United Kingdom}

\author{Alexander Welbourne}
\thanks{These authors contributed equally to this work.}
\affiliation{Department of Materials Science and Engineering, University of Sheffield, Sheffield, S1 3JD,  United Kingdom}

\author{Stephan J. Kyle}
\affiliation{Department of Materials Science and Engineering, University of Sheffield, Sheffield, S1 3JD,  United Kingdom}

\author{Paul W. Fry}
\affiliation{Department of Electronic and Electrical Engineering, University of Sheffield, Sheffield, S1 3JD,  United Kingdom}

\author{Dan A. Allwood}
\affiliation{Department of Materials Science and Engineering, University of Sheffield, Sheffield, S1 3JD,  United Kingdom}

\author{Thomas J. Hayward}
\affiliation{Department of Materials Science and Engineering, University of Sheffield, Sheffield, S1 3JD,  United Kingdom}

\author{Eleni Vasilaki}
\affiliation{Department of Computer Science, University of Sheffield, Sheffield, S1 4DP, United Kingdom}%

\begin{abstract}
The impressive performance of artificial neural networks has come at the cost of high energy usage and CO$_2$ emissions. Unconventional computing architectures, with magnetic systems as a candidate, have potential as alternative energy-efficient hardware, but, still face challenges, such as stochastic behaviour, in implementation. Here, we present a methodology for exploiting the traditionally detrimental stochastic effects in magnetic domain-wall motion in nanowires. We demonstrate functional binary stochastic synapses alongside a gradient learning rule that allows their training with applicability to a range of stochastic systems. The rule, utilising the mean and variance of the neuronal output distribution, finds a trade-off between synaptic stochasticity and energy efficiency depending on the number of measurements of each synapse. For single measurements, the rule results in binary synapses with minimal stochasticity, sacrificing potential performance for robustness. For multiple measurements, synaptic distributions are broad, approximating better-performing continuous synapses. This observation allows us to choose design principles depending on the desired performance and the device's operational speed and energy cost. We verify performance on physical hardware, showing it is comparable to a standard neural network.
\end{abstract}


\date{3rd March 2023}




\maketitle

\section*{Introduction}\label{sec1}

The meteoric rise of artificial intelligence (AI) as a part of modern life has brought many advantages. However, as AI programs become increasingly more complex, their energy footprint becomes larger  \cite{Thompson2021,SabryAly2019N3XT}, with the training of one of today's state-of-the-art natural language processing models now requiring similar energy consumption to the childhood of an average American citizen~\cite{Strubell2019_NLPenergy}. Several non-traditional computing architectures aim to reduce this energy cost, including non-CMOS technologies  \cite{ziegler2020novel,niemierNanomagnetLogicProgress2011,finocchio2021promise,Grollier2020NeuroSpintron}. However, competitive performance with non-CMOS technologies requires overcoming the latent advantage of years of development in CMOS.   

In biological neural networks, synapses are considered all-or-none or graded and non-deterministic, unlike the fully analogue synapses modelled in artificial networks \cite{petersen1998all}. Inspired by biology, several approaches have considered networks with binary synapses and neurons, with the view that binary operations are simpler to compute and thus lower energy \cite{simonsReviewBinarizedNeural2019, laborieuxLowPowerInMemory2020, yu2013stochastic, penkovskyInMemoryResistiveRAM2020}. However, while these binarised neural networks are more robust to noise, they suffer from lower performance than analogue versions. In contrast, networks with stochastic synapses provide sampling mechanisms for probabilistic models \cite{neftci2016stochastic} and can rival analogue networks at the expense of long sampling times \cite{daniels_energy-efficient_2020,liHEIFHighlyEfficient2019,shao2021implementation,muthappaHardwarebasedFastRealtime2020,nisarImplementationEfficientMagnetic2020, hirtzlin2019stochastic}. Adapted training methods are required to provide higher performance for a lower number of samples, while implementations require hardware that can natively (with low energy cost) provide the stochasticity required.
Magnetic architectures are one possible route for unconventional computing. They have long promised a role in computing logic following the strong interest in the field stemming from the data storage market \cite{Baibich1988,binaschEnhancedMagnetoresistanceLayered1989,allwood2005magnetic,Mccray2009,parkinMagneticDomainwallRacetrack2008,Lavrijsen2013Magnetic-ratche,Fernandez-Pacheco2016b,Grollier2020NeuroSpintron,finocchio2021promise}. The non-volatility of magnetic elements naturally allows for the data storage, while ultra-low-power control mechanisms, such as spin-polarised currents or applied strain \cite{emori2013current, hu2011high} offer routes towards energy-efficient logic-in-memory computing. Ongoing developments have shown how to manipulate magnetic domains to both move data and process it \cite{allwood2005magnetic, parkin2008magnetic, Sanz-Hernandez2017a,luo2020current}. However, magnetic domain wall logic is limited by stochastic effects, particularly when compared to the low error tolerance environment of CMOS computing  \cite{hayward2015intrinsic, kumarDomainWallMotion2019}.

Here we propose a methodology where, rather than seeking to eliminate stochastic effects, they become a crucial part of our computing architecture. As a proof of concept, we demonstrate how a nanowire is usable as a stochastic magnetic synapse able to perform handwritten digit recognition using multiplexing of one of the hardware synapses. 

We have developed a learning rule that can effectively train artificial neural networks made of such ``noisy'' synapses by considering the synaptic distribution. Suppose we allow a single measurement to identify the state of the synapse. In that case, the learning rule will adjust its parameter, i.e. the field at which the wall is propagated, to reduce the synaptic stochasticity. If we allow multiple measurements, the gradient rule will find parameters that allow for a broad synaptic distribution, mimicking a continuous synapse and improving performance. Without the stochasticity, the operation would be limited to binary operations, which lack the resolution power of analogue synapses. With stochasticity, we have a flexible system tunable between quick-run-time approximation and long-run-time performance. Our learning rule provides efficient network training despite the high or variable noise environment and differs from other stochastic neural network computing schemes that employ mean-field-based learning rules~\cite{ hirtzlin2019stochastic,daniels_energy-efficient_2020,shao2021implementation}. Here, the inclusion of the network variance allows the training to find better solutions in low sampling regimes, providing a trade-off between operational speed/energy cost and test accuracy.

We have verified the model performance experimentally by transferring the trained weights to a network utilising such a hardware synapse, with excellent agreement between the experimental performance and that of a simulated network. Our observations allow for a design framework where we can identify the number of required measurements (and hence energy requirements) for a given desired accuracy and vice versa.

This work opens up the prospect of utilising the low-energy-cost benefits of spintronic-based logic  \cite{lambsonExploringThermodynamicLimits2011,niemierNanomagnetLogicProgress2011, Grollier2020NeuroSpintron,finocchio2021promise}. In particular, it enables the use of domain wall-based nanowire devices  \cite{hayashiCurrentControlledMagneticDomainWall2008,parkinMagneticDomainwallRacetrack2008,Yang2015a,luo2020current} whilst transforming the hitherto hindrance of noisy operation \cite{hayward2015intrinsic,kumarDomainWallMotion2019} into the basis of a high-performance stochastic machine learning paradigm.

\section*{Results}\label{sec2}

\subsection*{Hardware stochastic synapse}

\begin{figure*}
\centering
\includegraphics[width=\linewidth]{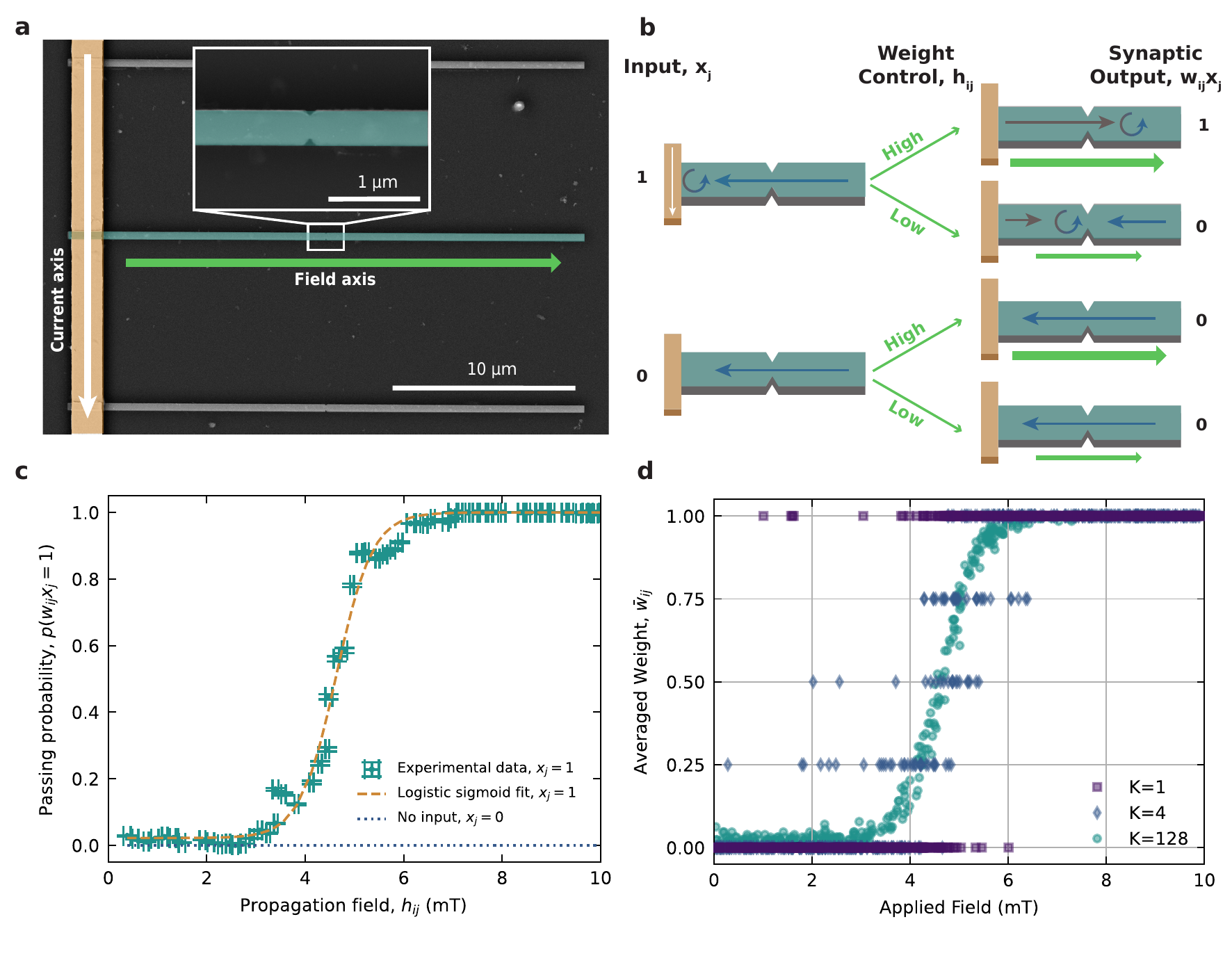}
\caption{{\bf Characterisation of notched permalloy nanowire (NW) as a stochastic synapse.} {\bf a,} False-coloured SEM image of the permalloy NW (blue) and current injection line (orange). The inset shows detail of the artificial notch. The field (green) and current  (white) axes are marked.  {\bf b,} Schematic of the operating principle of the stochastic synapse. The current line allows input ($x_j$) of 1 (current pulse, DW injected) or 0 (no pulse, no DW). Field inline with the NW drives (if present) the DW through the system: high fields pass the DW through the notch and produce an output of 1, low fields result in the notch blocking the DW and an output of 0. Intermediary fields (not shown) provide intermediate probabilities of passing the notch. {\bf c,} Experimentally measured probability of an injected domain wall passing the notch. Tuning the propagation field can control this probability across the whole range in a logistic sigmoid-like fashion. Points are averages of 1000 samples, x error bars represent precision in choice of propagation field, and y error bars are given $p(1-p)/\sqrt{1000}$.  The logistic sigmoid fit is given in methods. The nucleation field with no input (no injection, $x_j = 0$) is $(10.74 \pm 0.07)$ mT. Therefore, below \SI{10}{\milli\tesla} the passing probability for no input is zero. {\bf d,} Average synaptic weight, as defined in eq.~\ref{eq:mean_syn}. Depending on the number of samples, i.e. repetitions of the operation in {\bf b}, the effective synaptic weight varies from purely binary (one repetition/sample, $K=1$) to almost continuous ($K=128$ samples).}
\label{fig:DW_pinning}
\end{figure*}

Our proposed elementary computation unit is a binary stochastic synapse based on a ferromagnetic nanowire with two favourable magnetic orientations. The transitions between regions of differing magnetisation orientation are known as domain walls (DWs). While different forms of DWs exist, here they form a `vortex' pattern with a cyclical magnetisation texture. Our synapse was a \SI{400}{nm} wide, \SI{54}{\nano\meter} thick permalloy nanowire with notches patterned halfway along its length to create an artificial defect site. Figure \ref{fig:DW_pinning}.a shows an SEM image of the system, with the inset enlarging the notch. DWs were nucleated at the left-hand side of the wire (false-coloured blue) by applying a voltage pulse across a gold current line (false-coloured orange).

The operation of this system as a stochastic synapse is described schematically in figure \ref{fig:DW_pinning}.b. A vortex DW \cite{nakataniHeadtoheadDomainWalls2005a} can be injected into the wire by applying a current pulse in the line. This corresponds to presenting the synapse with an input of 1, while no DW injection corresponds to an input of 0. An applied magnetic field is used to propagate the DW along the length of the wire. If the propagation field is sufficiently high, the DW does not pin at the defect site and can pass to the end of the wire, resulting in an output of 1. If the propagation field is low, the DW is pinned at the notch, resulting in an output of 0. For intermediate values of the field, the behaviour becomes stochastic but with a well defined pinning probability. We can consider the field control as controlling the weight in a binary synapse with detecting a DW on the right hand side of the nanowire as the output of the synapse. 

As the propagation field is tuned, the probability of the DW passing changes. Figure \ref{fig:DW_pinning}.c shows this passing probability, as measured using the focused Magneto-Optical Kerr effect (FMOKE), as a function of the propagation field. The probability of passing behaves in a sigmoid-like manner, and the orange dashed line shows a fit using a logistic sigmoid function $f(h_{ij})$ (see methods).

Therefore, a binary stochastic synapse is determined by
\begin{equation}
	w_{ij} = \begin{cases} 1 & \text{with probability  } f(h_{ij}) \\
    0 & \text{otherwise,}
    \end{cases}
    \label{eq:weights}
\end{equation}
where $f(h_{ij})$ is the DW passing probability function, $h_{ij}$ is the propagation field for the synapse connecting input neuron $j$ with output neuron $i$. Through this definition our synapses are purely excitatory, which corresponds to the physical representation of a magnetic DW being pinned or not, rather than the complementary binary scheme with values $\{-1,1\}$, which is not naturally represented by the physical system. 

Compared to binary synapses, neural networks with analogue or graded synapses tend to perform better due to the wider range of states  \cite{satel2009binary, dubreuil2014memory}. Here, we adopt a scheme similar to that of stochastic computing, where the average of a series of binary measurements or samples are used to represent a value.
Thus, we allow for $K \geq 1$  measurements to identify the state of a synapse and denote the equivalent mean weight as

\begin{equation}\label{eq:mean_syn}
    \bar{w}_{ij} = \frac{1}{K} \sum_k w_{ij}^{(k)},
\end{equation}
where $K$ is the total number of samples taken and the superscript $(k)$ indicates the individual sampling of the synaptic weights as per eqn. \ref{eq:weights}. The mean synapse has $1+K$ states, e.g. for $K=1$ the two states will be 0 and 1, while for $K=2$ the states will be 0, 0.5, and 1. It follows that for $K \rightarrow \infty$, $\bar{w}_{ij}$ will be equivalent to a sigmoidally-shaped continuous synapse, bounded between 0 and 1. An example demonstrating the average weight as a function of the number of samples can be seen in figure \ref{fig:DW_pinning}.d, where we plot 
eq.~\ref{eq:mean_syn} for $K=1$ (purple squares), 4 (blue diamonds) and 128 (green circles). Each example is calculated by sampling $w_{ij}$ the desired number of times with a fixed $h_{ij}$ that was selected randomly. In each case only discrete levels are available but when $K=128$ the sampling is sufficient to provide an almost continuous representation.

In this way, our proposed binary stochastic synapse can be used to construct neural networks that will approach a bounded analogue network when multiple samples are taken. Physically, this is achieved by repeated operation of the hardware devices to accumulate the average values.

\subsection*{Stochastic network}

\begin{figure*}[htbp!]
\centering
\includegraphics[width=\linewidth]{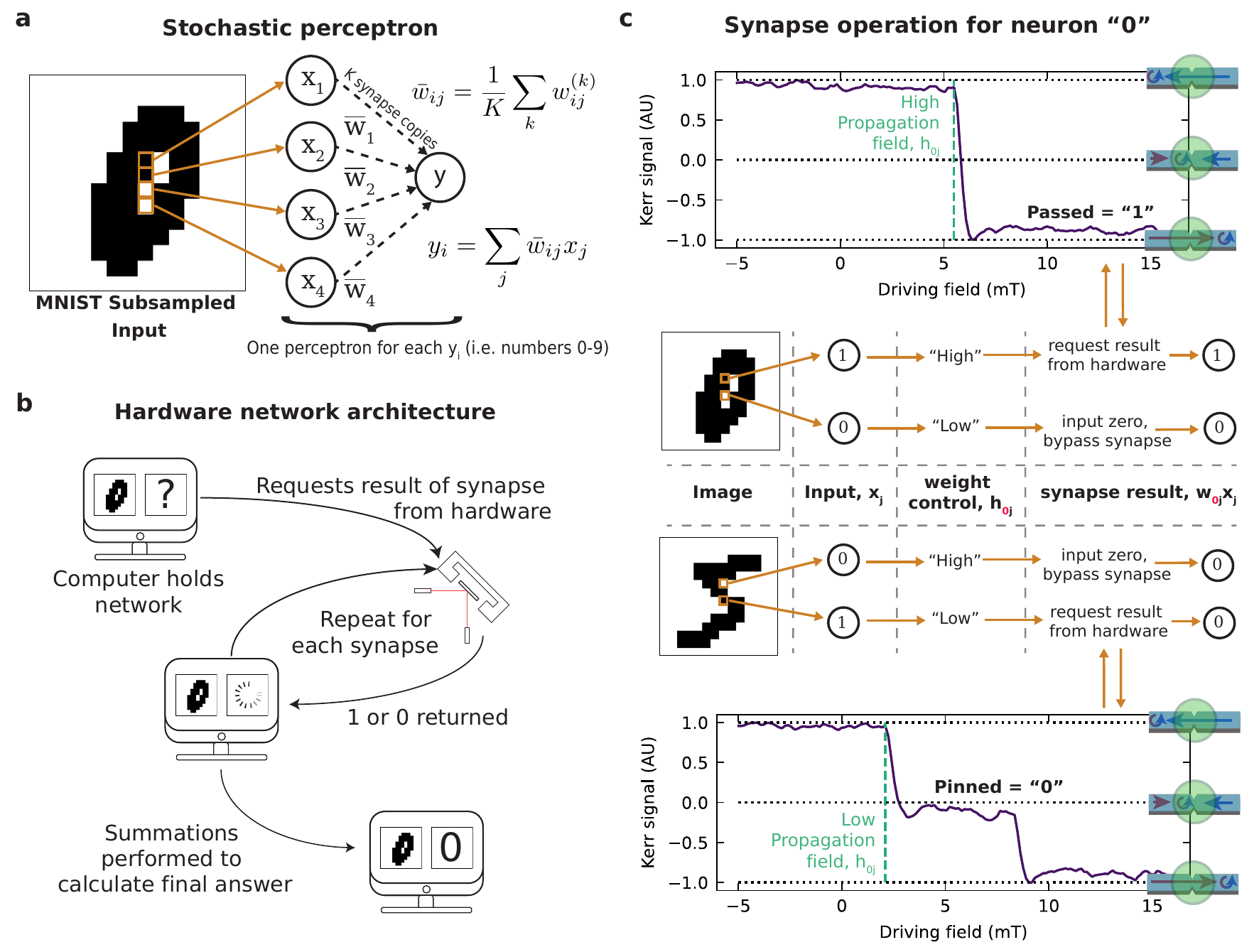}
\caption{{\bf Stochastic network operation.} \textbf{a}, Sketch of the stochastic perceptron. Each input value from an image is fed via a mean weight to the neurons for each class. Here, the weighted inputs are summed to give the neuron's activity (as in eq.~\ref{eq:neuron}). In the case of the MNIST task, there is a neuron for each of the 10 classes (numbers ``0'' to ``9''; $y_0$ to $ y_9$). When trained, the neuron for the class corresponding to the correct input, here $y_0$, should have the highest activity. Each mean weight in our network ($\bar{w}_{ij}$) is the average of multiple measurements ($K \geq 1$) of the output of a synapse with individual weight $w_{ij}$ set by its trained propagation field (see eqs.~\ref{eq:weights}~\&~\ref{eq:mean_syn}). A clear distinction should be made here from traditional neural networks that these weights are stochastic and will vary for each run of the network. The individual weights take the value ``1'' with the probability $f(h_{ij})$ (DW passing probability, as characterised in figure~\ref{fig:DW_pinning}.c) or ``0'' otherwise. The mean weights, therefore, take values from the distributions shown in figure~\ref{fig:DW_pinning}.d. 
\textbf{b}, The architecture of the hardware network. For the purpose of demonstrating successful performance, only the stochastic synapses are run directly on the hardware. The perceptrons are stored on a computer, which requests results (``1'' or ``0'') from the magnetic stochastic synapses for a given synaptic parameter (trained propagation field, $h_{ij}$). After this is repeated for each synapse, summations are performed to predict the correct class of the input. \textbf{c}, Idealised operation of single synapses \textit{in materia} for the neuron $y_0$. The data path is shown for two inputs, or pixels, for the case of a correct image for the class (``0'') and an incorrect image (``5''). The value of the weight control, the propagation field $h_{ij}$, is expected to be correlated with pixels in images from the correct class: where the pixels are ``on'' for correct images, high values of the weight control are expected; when ``off'', low values. If the input pixel value is ``0'', the synapse is bypassed as the result is ``0'' by construction. However, if it is ``1'', a result is requested from the hardware using the corresponding weight control. As shown in the top graph, high propagation fields result in the DW directly passing the notch (only a single step is seen) which is interpreted as an output of ``1''. As in the lower graph, low propagation fields result in a two step procedure where the DW initially pins at the notch before depinning at a higher driving field. This is interpreted as a ``0''. In practice, the results from the synapses will vary stochastically reflecting the passing probability $f(h_{ij})$. }\label{fig:architecture}
\end{figure*}

We embed these synapses in an artificial neural network where the output of neuron $i$ is given by
\begin{equation}
   y_i = \sum_j  \bar{w}_{ij} x_j,
   \label{eq:neuron}
\end{equation}
where $j$ is an index over the input dimension. 

We trained the network as a classifier for a problem of $C$ classes with $C$ independent neurons (perceptrons), where each neuron represented one class. This task was based on the well-known MNIST dataset but with each image downsampled to give images with a shape of 14 by 14 pixels instead of the standard 28 by 28. This was necessary to reduce the time of the operation when running on the prototype experimental hardware (see methods). In figure \ref{fig:architecture}.a we depict the perceptron that corresponds to class ``0". If we present to the neuron a representative of its corresponding class (in this case an image of the digit ``0"), the neuron should produce a high activity for recognising the input as zero.

The experimental process is shown in figure \ref{fig:architecture}.b. For ease of demonstration, only a single hardware synapse is used, with operations serialise in time. Potential devices would have multiple synapses running in parallel with a summation performed during the measurement. The perceptron parameters are stored on a computer, which sends the input and synaptic parameter to the external hardware synapse and requests the result. The process is repeated until $K$ samples per synapse (see eq. \ref{eq:mean_syn}) are collected. Summation of the results takes place on the computer with an additional bias term applied. To avoid redundant measurements, pixels corresponding to inputs of ``0" (white pixels in our example image) were omitted, since the output is deterministically ``0" by design. A synapse receiving a black pixel ($x_j = 1$) will produce ``1'' if the field is set at a high value or ``0'' if the field is set at a low value, see figure \ref{fig:architecture}.c. Intermediate field values will produce outputs that vary scholastically, reflecting the passing probability $f(h_{ij})$.


\subsection*{Analysis of the stochastic learning rule}
We now sketch the derivation of the learning rule that we apply to the synapses of the neural network. Each synapse $w^{(k)}_{ij}$ is an independent sample from a Bernoulli distribution, and therefore the sum of these samples will follow a Poisson-Binomial distribution. The mean, $\mu_i$, and variance, $\sigma_i^2$, for each output neuron (calculated by eq. \ref{eq:neuron}) are given by:
\begin{align}
    \mu_i &= \sum_j f(h_{ij}) x_j, \label{eq:mean} \\
    \sigma^2_i &= \frac{1}{K}\sum_j f(h_{ij}) x_j \left[ 1 - f(h_{ij}) x_j\right]. \label{eq:var}
\end{align}
For a detailed calculation of these values see the supplementary material.

Since the number of inputs and the sampling process means this sum will be over a large number of events, the Poisson-Binomial distribution can be approximated as a Gaussian \cite{hong2013computing}. Using this approximation, the neuronal output can be re-parameterised so that the stochasticity is only in a term with no dependence on the trainable parameters. In this way, we write

\begin{equation} \label{eq:gaussian_approximation}
    \tilde{y}_i = \mu_i + \sigma_i \xi_i,
\end{equation}
where $\tilde{y}_i$ denotes the approximation of neuronal output $y_i$ and $\xi_i$ is a sample from a Gaussian distribution with zero mean and unit variance. 

If we assume that we are in a supervised learning framework and that $E$ is the error function we would like to minimise (e.g. square mean error or cross-entropy), then $E$ is a function of the pattern $p$ we present to the network, which defines the desirable output target. $E$ is also a function of the output neurons, represented by vector $y$, which also depends on $p$. The learning rule will update the values of the applied field to each synapse $h_{ij}$ by $\Delta h_{ij}$ according to the following ``online'' gradient rule:

\begin{align}
	\Delta h_{ij} & =-\eta \frac{\partial E(\tilde{y})}{\partial \tilde{y}_i}\frac{\partial \tilde{y}_i}{\partial h_{ij}} \\
    & = - \eta \frac{\partial E(\tilde{y})}{\partial \tilde{y}_i} \left( \frac{\partial \mu_i}{\partial h_{ij}} + \frac{\partial \sigma_i}{\partial h_{ij}} \xi_i \right),
\end{align}
where $\eta$ is a small positive number representing the learning rate. We calculate the derivative of $\frac{\partial \tilde{y}_i}{\partial h_{ij}}$ from eq. \ref{eq:gaussian_approximation}, \ref{eq:mean}, \ref{eq:var}. We also calculate the value $\xi_i$ using eq. \ref{eq:gaussian_approximation}, computing $\mu_i$ and $\sigma_i$ from  eq. \ref{eq:mean} and \ref{eq:var} and $\tilde{y}_i $ from eq. \ref{eq:neuron} (setting $\tilde{y}_i=y_i$). It follows that for $K \rightarrow \infty$, $\sigma \rightarrow 0$ and we obtain a ``mean-field'' gradient rule that takes into account the mean but not the variance of the output neurons.

\begin{figure*}[htbp]
\centering
\includegraphics[width=\linewidth]{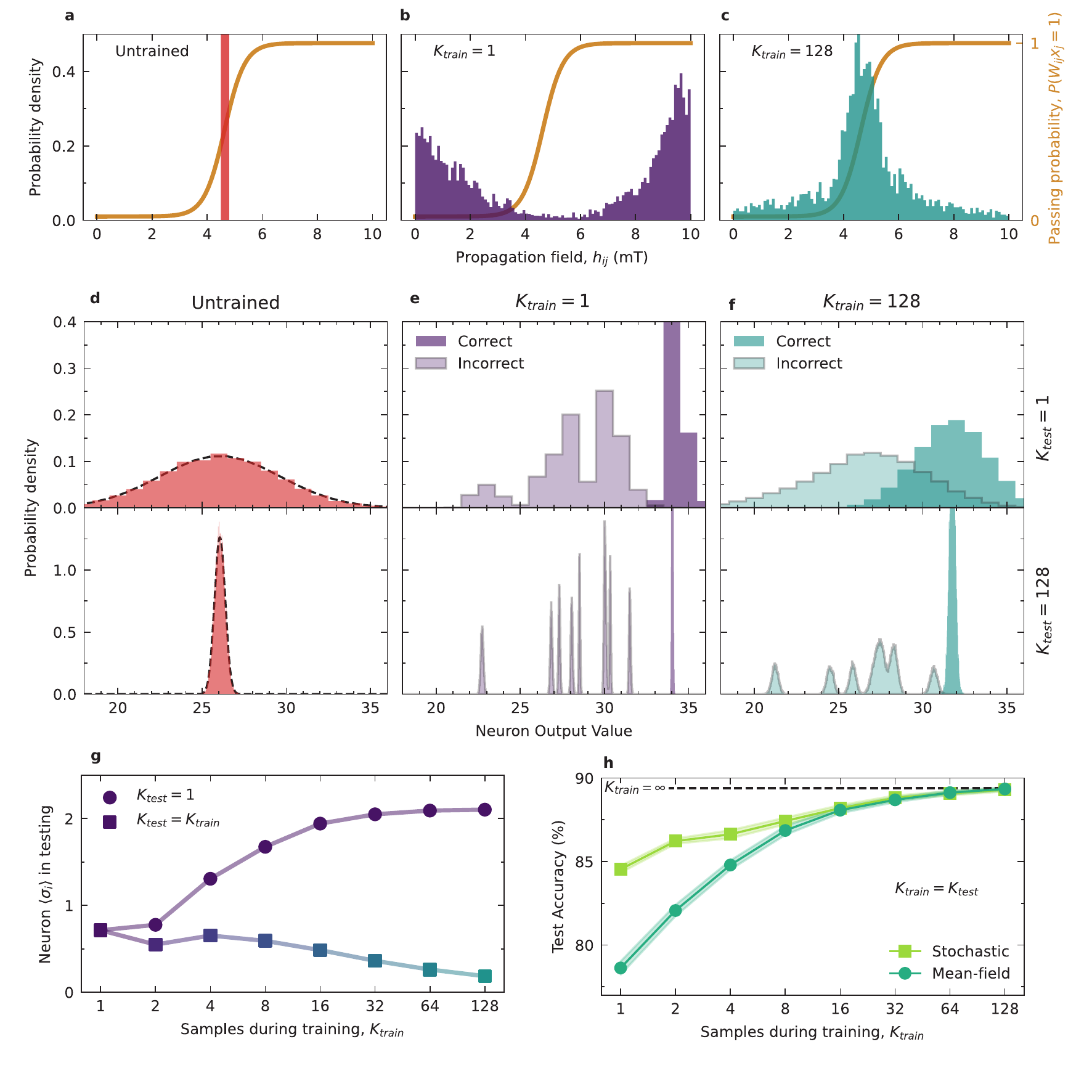}
\caption{{\bf Analysis of the stochastic learning rule.} {\bf a-c}, Probability density histograms of synaptic magnetic field parameters over 5 independent models. {\bf a} shows the distribution before training, where all the fields are initialised so that the passing probability (shown in orange, right hand axis) is 0.5. {\bf b} and {\bf c} show the distributions when trained using 1 or 128 samples respectively. With 1 sample, the distribution is bimodal with peaks at fields with probabilities close to 0 or 1. While when training with 128 samples, the distribution is focused on the central region of the passing probability function. {\bf d-f}, Distribution of the neuron values $\mathbf{y}$ when an image of a zero is shown 10,000 times independently for neurons either identifying the correct (output 0 in this example) or incorrect (outputs 1-9) classification. In {\bf d} the model is untrained so all outputs have the same distribution, while in {\bf e} and {\bf f} the distribution is split into the correct output neuron and the incorrect output neurons when training with 1 and 128 samples respectively. The top row shows $K_\mathrm{test}=1$ while the bottom row shows $K_\mathrm{test}=128$. Using more samples during testing reduces the variance and therefore the chance of mis-classification. {\bf g}, The standard deviation averaged over all the neurons when increasing number of samples are used in training, with $K_\mathrm{test}=1$ (circles) and $K_\mathrm{test} = K_\mathrm{train}$ (squares). This summarises the conclusions from the distribution plots in {\bf d-f}. More samples during training allows the standard deviation for a single sample to increase as the standard deviation over all samples is reduced. However, testing with $K_\mathrm{test} < K_\mathrm{train} $ results in an increased overall standard deviation. {\bf h}, Accuracy on the test set against number of samples during training when using the stochastic (dark green circles) or the mean-field (light green squares) learning rules. The points show the accuracy averaged over 5 independently trained models, while the shaded region indicates 1 standard deviation. The stochastic learning rule maintains a higher test accuracy when the number of samples is low.}
\label{fig:training}
\end{figure*}

We have tested the performance of this rule on the downsampled MNIST dataset. During training, the number of repeats (samples) $K$ is set as a parameter of the network, which we define as $K_\mathrm{train}$, and as such modifies how the training progresses. The variance of the output has an important effect on the classification procedure; if the variance is high then mis-classification will be more likely, especially in classes that have similar mean values for each neuron. Therefore, during supervised training the network aims to minimise this variance. When K is low, this happens through changing the weights, controlled through the magnetic fields $h_{ij}$, so that the probabilities are close to either 1 or 0 (high or low applied field), as this minimises the single sample variance in eqn. \ref{eq:var}. This leads to a solution that is almost a deterministic binary network. However, if K is large then the variance is reduced by the factor $1/K$ and therefore the system can tolerate higher synaptic variance than in the case of $K=1$. Thus, a pseudo-analogue solution can be found.

Figure \ref{fig:training} describes the effect of the learning rule on the network synapses. We plot the distribution of the propagation fields, $h_{ij}$, over all the neurons from 5 independent models before training (figure \ref{fig:training}.a), after training with $K_\mathrm{train}=1$ sample (figure \ref{fig:training}.b) and after training with $K_\mathrm{train}=128$ samples (figure \ref{fig:training}.c). The final distributions confirms the theoretical expectation that $K_\mathrm{train}=1$ leads to a binary network (low variance) while $K_\mathrm{train}=128$ approximates a standard perceptron with a continuous distribution of synaptic weight (high variance).

In figure \ref{fig:training}.e-f we show the distributions of the neuronal output when presented with the same image repeatedly for the three training cases above and find that the neuronal distribution reflects the synaptic distribution. We now consider the case where during testing a different number of samples are drawn when calculating eqn. \ref{eq:mean_syn}, which we define as $K_\mathrm{test}$. The top row shows the distribution when $K_\mathrm{test}=1$, while the bottom row shows $K_\mathrm{test}=128$. The untrained neuron values exhibit a Gaussian distribution across all data samples, with fields initialised to give the largest possible variance; see figure \ref{fig:training}.d. After training, with $K_\mathrm{train}=1$ or $K_\mathrm{train}=128$ and $K_\mathrm{test} = K_\mathrm{train}$ the distributions of the correct and incorrect class neurons minimally overlap. However, if we test the network with $K_\mathrm{test}=1$ after we train it with $K_\mathrm{train}=128$ there is a rather significant overlap (\ref{fig:training}.f, upper panel) suggesting a high probability of miss-classification. In all cases, when testing with $K_\mathrm{test}=128$ (bottom row) the variance is reduced $1/128$ as given in eqn. \ref{eq:var} and allows for better resolution of the mean values. \new{In the case $K_\mathrm{train}=128$, the learning rule has exploited this additional sampling and variance reduction by better utilising a continuous range of 
weights to boost performance. However, when $K_\mathrm{test}<128$ (as in \ref{fig:training}.f, upper panel), the increase in variance decreases the probability of correct classification. In the other training case ($K_\mathrm{train}=1$, \ref{fig:training}.e, upper panel), the learning rule adapts the weights to find a low variance, almost deterministic binary, solution. Further sampling during testing (\ref{fig:training}.e, lower panel) reduces this variance further, as expected, but doesn't significantly change the overlap as it has already been optimised for the lower sampling regime. As we will show, this leads to higher performance when test sampling ($K_\mathrm{test}$) is small, but capped high performance when test sampling is allowed to rise, in contrast to the large $K_\mathrm{train}$ case.}

Figure \ref{fig:training}.g compares the average variance during testing with $K_\mathrm{test}=1$ samples (circles) and $K_\mathrm{test}=K_\mathrm{train}$ samples (squares) as a function of the number of samples used during training, $K_\mathrm{train}$. As discussed before, when training with 1 sample the variance is kept low by having passing probabilities close to 0 or 1. However, when more samples are used during training, the variances for a single sample can increase as the variance of the averaged samples decreases.

This behaviour of minimising the variance to reduce miss-classification arises due to the variance term in the ``stochastic'' learning rule. Other rules that only consider the mean term  \cite{hirtzlin2019stochastic, daniels_energy-efficient_2020} cannot find these deterministic solutions when using a stochastic network. Figure \ref{fig:training}.h shows the test accuracy with $K_\mathrm{test}=K_\mathrm{train}$ as a function of $K_\mathrm{train}$ samples for our stochastic learning rule (squares) vs the mean field rule (circles) averaged over five independently trained models. For both rules, increasing the number of samples leads to an improvement in the test accuracy as more levels are possible for the synapse averages (see fig. \ref{fig:DW_pinning}.d). However, in all cases, the stochastic learning rule out performs the mean-field rule, with convergence when a large number of samples ($K \geq 8$) is used for training and testing. The dashed line shows the performance for a fully mean-field network, where effectively an infinite number of samples are taken (i.e continuous but bounded synapses), and represents the best possible accuracy for such a network given the task. 

\subsection*{Hardware and operational principles.}

\begin{figure*}[tbp]
\centering
\includegraphics[width=1\textwidth]{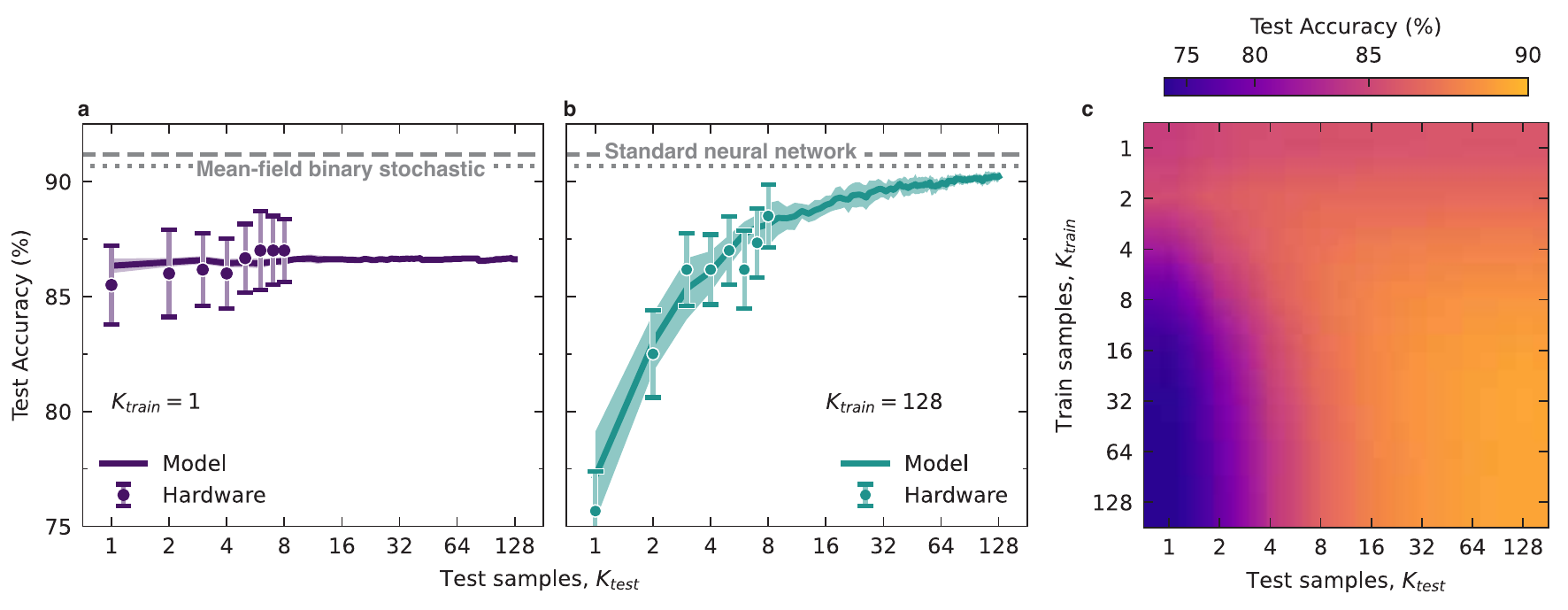}
\caption{{\bf Hardware verification and choice of sampling.} {\bf a,b} Comparison of the testing accuracy computed using either the physical hardware (points) or from simulation (curves). The test data set is restricted to the first 600 images with an approximately equal balance of digits. In both, training was done using the model, and hardware testing was limited to eight samples due to throughput limitations of the prototype device. {\bf a} shows the accuracy when the network was trained with only 1 sample while {\bf b} was trained with 128 samples. As before, training with 1 sample reaches an almost deterministic solution, so repeated sampling during testing does not improve the accuracy. Training with 128 shows an increase in accuracy as more samples are used during testing, reaching higher peak performance (albeit with a lower initial base). The dashed line shows the performance on a standard neural network and the dotted-dashed is for a full mean-field binary stochastic network. In both cases the hardware performance shows excellent agreement with the model calculations. The model accuracy is averaged over 5 independent tests with the same trained weights, with the shaded area showing 1 standard deviation. This can be taken to represent the variability in performance for a given task due to the inherent stochasticity of the network. Naturally, it decreases as the number of test samples increases and is lower for the, more deterministic, $K_\textrm{train} = 1$ case. The hardware accuracy is from a single run over the 600 images, so the error bars show the standard error of the estimation of the accuracy over the mini-batches. {\bf c} Test accuracy (as measured with the model) over different combinations of training and testing sampling for the sub-sampled MNIST task. The data is bi-linearly interpolated, which can be considered as averaging over fractions of the data set with different sampling rates. In general, testing with more samples increases accuracy, but, this is limited when $K_\textrm{test} > K_\textrm{train}$. In particular, in the  $K_\textrm{train} =  1$ case, further sampling provides little improvement due to the deterministic weight distributions. Training with 2 samples is better in all test cases than when training with 1, but best overall accuracy is when 128 samples are used in both training and testing. Data such as this provide a guide to choosing training and testing samples depending on desired accuracy and operation times for a given task.}
\label{fig:device}
\end{figure*}

We now proceed to demonstrate our neural networks working on physical hardware and not only within simulation. Figures \ref{fig:device}.a and b shows the test accuracy computed when the synaptic operation has been simulated (lines) and processed using the hardware (points) for models trained with either a, $K_\mathrm{train}=1$ or b, $K_\mathrm{train}=128$ and tested with increasing numbers of samples ($K_\mathrm{test}$). Due to the throughput of our prototype device, we only demonstrate experimental results up to $K_\mathrm{test} = 8$.

The simulation and hardware results show excellent agreement and highlight different behaviours in models trained with different sampling levels. In the case trained on one repeat, the network is deterministic and as such the accuracy does not significantly improve when we average over more samples during testing. On the other hand, while the model trained with 128 repeats shows a lower performance with only one testing repeat, the accuracy improves as we increase the number of samples during testing. This arises from the increased stochasticity at low sampling levels and resultant increased precision at high sampling levels. This behaviour is corroborated by the corresponding neuronal distributions (\ref{fig:training}.e) and (\ref{fig:training}.f), which show that the neuronal variance when training with one sample and testing with one sample is much lower than in the case of training with $K=128$ samples and testing with one sample. It is akin to majority voting, where classifiers have to be diverse to improve performance (see  \cite{vouros2018generalised} and references therein). Here,  performance increase increases with increasing  $K_\mathrm{test}$ (number of voters) when the neuronal distribution has a high variance.



Figure \ref{fig:device}.{\bf c} allows further interrogation of the majority voting behaviour. It presents (using the now verified simulation model) a colour plot of the test accuracy as a function of the number of training and test samples. This variation in performance when testing using a different number of repeats raises an essential trade-off in speed vs accuracy. To a first approach, the results follow the behaviour of stochastic computing: fast approximation with increasing accuracy over time if required. This trend is matched on average with the extra repeats, implying an extra time and energy cost to accumulate the samples, but providing a boost in accuracy. However by utilising our learning rule's ability to enable low sampling deterministic solutions we can outperform the naive stochastic computing reasoning in the low sampling limit (as also seen in figure~\ref{fig:training}.h). If fixing $K_\mathrm{train}$, this leads to a competition between low repeat performance and ultimate high repeat accuracy, i.e if a model is trained on a high number of repeats, but uses a low number of repeats during testing (inference) time, then the accuracy will be sub-optimal. Similarly, if the model is trained on a low number of repeats, but tested on many, the ultimate accuracy suffers. One possibility is to always tie $K_\mathrm{test} = K_\mathrm{train}$, but this requires multiple trained weights. It is, therefore, constructive to utilise data such as Figure \ref{fig:device}.{\bf c} as a guide on training and testing the synapse depending on the desired accuracies and operational times. Whilst maintaining the simplicity of a single set of trained weights (fixing $K_\textrm{train}$), a horizontal range of testing values can be chosen to achieve the desired accuracies and energy cost envelope.

Analysing the performance over the space of training vs test repeats for the MNIST task, we find that in most cases, testing with a similar number of repeats to the training performs well. A significant outlier to this was that training with two samples consistently outperformed training with one across all levels of testing, including testing with one sample. We attribute this to the smaller step sizes in the parameter space with two samples compared to one, which allows for a better solution while the variance is still very low and remains small when testing with one sample.

\section*{Discussion}\label{sec12}

\new{Neuromorphic devices are a promising route to developing low-energy-cost machine learning systems, seeking to overcome one of the chief drawbacks of traditional neural networks. Stochastic, binary neural networks have shown promise in this regard due to their reduced energy cost and simple implementation \cite{simonsReviewBinarizedNeural2019, laborieuxLowPowerInMemory2020, yu2013stochastic, penkovskyInMemoryResistiveRAM2020,neftci2016stochastic}. Multiple sampling of these networks allows their performance to rival analogue networks \cite{daniels_energy-efficient_2020,liHEIFHighlyEfficient2019,shao2021implementation,muthappaHardwarebasedFastRealtime2020,nisarImplementationEfficientMagnetic2020, hirtzlin2019stochastic}. Outstanding problems, however, have been providing training rules to achieve high performance even at low sampling rates (where calculations can be performed faster and at less energy cost) and identifying hardware implementations that can natively provide the stochasticity required. We have developed a learning methodology for stochastic binary neural networks that we verify experimentally, using the behaviour of magnetic domain walls in nanowires as stochastic synapses. Stochasticity has traditionally been considered a limiting factor in nanomagnetic logic devices \cite{hayward2015intrinsic,kumarDomainWallMotion2019}, but here is a functional aspect that drives learning. We have shown performance of the hardware network comparable to a standard neural network and demonstrated high performance at low sampling thanks to the novel learning rule.}


Experimentally, we have observed that a DW injected into a nanowire with an artificial pinning site can be stochastically pinned and tuned by using an applied magnetic field. We have then demonstrated that this tunable stochastic pinning can create synapses for a neural network device. Due to the nature of the physical system, these synapses behave as binary stochastic synapses. Our fundamental ingredient for training such a network is a learning rule that considers the variance of the stochastic output of the network. This training method considers taking multiple samples ($K_\mathrm{train}$/$K_\mathrm{test}$) of the network output to compute a sample average and deviation. A low number of samples leads toward a predominantly deterministic binary solution and is fast to compute but has lower performance than a high number of samples that approximates a standard ``analogue'' network and require more time (and energy). This trade-off allows flexibility in designing the network based on the required performance or operating speed.

Key is that the learning rule developed here has allowed us to find a range of operating regimes because the stochastic part of the output is considered. Other binary stochastic computing approaches, such as Hirtzlin \emph{et al.} \cite{hirtzlin2019stochastic}, train using the expectation of the network (which we call mean-field and is equivalent to $K \rightarrow \infty$) and leads to a reduced accuracy when fewer samples are used during inference (testing). The Gaussian approximation was also used by Esser \emph{et al.} \cite{esser2015backpropagation} to train a network with binary stochastic synapses on the IBM TrueNorth neurosynaptic system but the contribution from the variance term is considered to be negligible. The contribution from the variance term in our rule allows for weights to be trained that operate better in the low sampling regime compared to the mean-field versions.

Other learning methods where the variance is taken in account stem from the Likelihood-Ratio framework \cite{williams1992simple, gu2015muprop, parmas2021unified}, which is related to policy gradient methods in reinforcement learning \cite{vasilaki2009spike}. While these methods consider the stochasticity of the neurons and synapse, they depend heavily on the choice of baseline values for the loss which require complex approximation methods. Additionally, the reparameterisation method applied here allows for a direct feedback of the error signal to the synaptic field parameters and fits within existing backpropagation-based learning methodologies.


Overall, the stochastic learning rule presented in this paper has shown tunability in both high and low sampling regimes and can be implemented simply within backpropagation-style codes. The ability, due to consideration of the variance of the output, to tune between low-sampling deterministic binary and high-sampling stochastic ``analogue like'' behaviour lends itself to the flexibility of our system between operational
speed/energy cost and test accuracy.


The magnetic DW synapse that we have demonstrated here is a proof of principle component and as such it important to look towards changes that would be necessary for a more ``production ready'' neuromorphic device. Optimised devices would likely look towards spin-torque driven domain wall motion \cite{luo2020current} alongside the use of local nanomagnetic elements to encode the weights.
It is also possible to envisage our learning methodology applied to networks built of alternative magnetic elements with similar stochastic properties, such as magnetic tunnel junctions, amongst others \cite{daniels_energy-efficient_2020, penkovskyInMemoryResistiveRAM2020, azam2020voltage,sanz-hernandezTunableStochasticityArtificial2021, misbaEnergyEfficientLearning2022}. Elsewhere, DW devices have been used as neurons \cite{hassan2018magnetic} or activation functions \cite{brigner2022domain} and magnetic elements in general have been demonstrated in a range of alternative low energy computation schemes \cite{borders2019integer,shao2021implementation, al2022energy, azam2020voltage, koo2020sbsnn} that exploit the stochasticity of magnetic devices.
Our fundamental element, the magnetic stochastic synapse, could fit within such paradigms where efficient production of random bits is key. It is important, however, to state that the key result here is demonstrating performance as run on experimental hardware, enabled by our stochastic learning rule. Further optimisation is a matter of future research and engineering development.


Whilst the single layer network demonstrated here can only solve linearly separable problems, it \new{can be extended in a number of ways. Retaining the single layer simplicity and looking towards an all magnetic architecture, it} has potential applications in the field of reservoir computing. In reservoir computing, a fixed reservoir performs a non-linear spatial-temporal transformation of an input sequence such that the output representation is linearly separable. The advantage of RC is that the reservoir transform can be offloaded to a physical system with appropriate properties and there has been considerable recent interest in developing magnetic (spintronics) based physical reservoir computing \cite{Grollier2020NeuroSpintron, Dawidek2021, Ababei2021, Welbourne2021,gartsideReconfigurableTrainingReservoir2022,vidamourQuantifyingComputationalCapability2022,vidamourReservoirComputingEmergent2022,allwoodPerspectivePhysicalReservoir2022,stenningAdaptiveProgrammableNetworks2022}. There is potential to connect our magnetic DW based neural network to these reservoirs to create a complete hardware reservoir computing system. 
\new{There is also the more traditional route of scaling our current approach towards multi-layer networks as the learning rule is compatible with back-propagation. An open research question in this avenue is whether the sampling procedure should apply at a local or global scale of the network. One approach is implementation of multi-layers using nanowire interconnects and logic gates, but if we look away from the limitation of all magnetic architectures, it is also possible to envisage hybrid magnetic-CMOS application specific integrated circuits (as in Ref. \onlinecite{hirtzlin2020digital}) that might provide a route to larger scale network hardware. However, details of these implementations are beyond the scope of this current work.}

In conclusion, we have \new{developed a training methodology for binary stochastic synapses that considers the network's stochasticity during learning and resampling of the stochastic output allows for a trade-off between device run time and desired accuracy. This approach has been demonstrated on a proof of concept magnetic domain wall-based stochastic synapse with excellent agreement between hardware and model during inference.}  

\FloatBarrier
\section*{Methods}

\subsection*{Device Fabrication}
The devices were fabricated using two-stage electron beam lithography with the CSAR-62 resist. Nanowires were deposited in the first stage using thermal evaporation of permalloy (Ni$_{81}$Fe$_{19}$) to a thickness of \SI{54}{\nano\meter} (base pressure, \SI{7E-7}{\milli\bar}; process pressure, \SI{\sim 5E-5}{\milli\bar}; rate, \SI{0.5}{\angstrom\per\second}). Current lines and connection pads were deposited in the second stage as Ti/Au (Nominally \SI{10}{\nano\meter}/ \SI{200}{\nano\meter} via thermal evaporation). Samples were electrically connected to PCB devices using silver DAG. 

\subsection*{Device operation}
The device operation procedes as in figure~\ref{fig:architecture}. An AVTECH pulse generator was used to apply 30 volt, 100-nanosecond pulses along the current line (resistance \SI{290}{\ohm}). An electromagnet was used to apply fields along the wire lengths. A National Instruments DAQ card was used to control timing between these two, with pulses being triggered at particular times during repeated sinusoidal field sequences. The field at which the pulse is triggered is the propagation field. On the fly calibration of timing enabled correction of any drift between the trigger and field sequence (due to heating) to  $\lesssim$ \SI{ 0.1}{\milli\tesla}.

A focused-MOKE magnetometer (spot size $\sim$ 5 micrometer) was used to measure the NW response. Hysteresis loops were obtained with the laser spot positioned over the notch. Single steps in the hysteresis loop indicate the domain wall passing the notch (an output of 1). Double steps indicate a two-stage pinning/depinning process (an output of 0). An algorithmic method allowed automated evaluation of each hysteresis loop. The number of peaks were calculated in the differentiated Kerr signal; if two peaks were present then the DW had been pinned. To eliminate false positives, the steps in the raw signal corresponding to the peaks were required to be greater than \SI{24}{\percent} of the total signal change. This was optimised experimentally to allow for peak detection even with a slightly off centre laser spot (unequal step sizes), but to minimises erroneous detections arising from noise. 

\subsection*{Domain passing probability}
The probability of a domain wall not being pinned by the artificial defect site was observed to have a sigmoid-like behaviour. A functional form of this probability was used to simulate magnetic stochastic synapses for computational training of the networks. We fitted this probability using

\begin{equation}\label{eq:sigmoidfit}
    f(h) = d + \frac{1-d}{1 + \exp(-\Delta (h-h_0))},
\end{equation}
where $d=0.0219$ is a finite passing probability at low field, $h_0 = \SI{4.63}{mT}$ is the field centre and $\Delta = \SI{2.73}{mT^{-1}}$ is the sigmoid width. We note that this exact form of the fitting function is not necessary for the stochastic learning rule used to train the network.

\subsection*{Stochastic learning rule}
For a network comprised of binary stochastic synapses the value of each neuron can be approximated by a Gaussian given by equation (6), where the mean and variance of each neurons is defined by equations (4) and (5) respectively. Using this approximation, a gradient based learning rule can be derived as the random variable no longer has a dependence on the model parameters. The parameters of this network are the magnetic fields which determine the passing probability of the synapse so a gradient descent update is given by
\begin{align}
    	\Delta h_{ij} &= -\eta \frac{\partial E(\tilde{y})}{\partial \tilde{y}_i}\frac{\partial \tilde{y}_i}{\partial h_{ij}} \\
    	& = - \eta \frac{\partial E(\tilde{y})}{\partial \tilde{y}_i} \left( \frac{\partial \mu_i}{\partial h_{ij}} + \frac{\partial \sigma_i}{\partial h_{ij}} \xi_i \right).
\end{align}
The gradient of the mean and variance with respect to the magnetic fields are
\begin{align}
\frac{\partial \mu_i}{\partial h_{ij}} & = f'(h_{ij}) x_j \\
\frac{\partial \sigma_i}{\partial h_{ij}} & = \frac{(1- 2f(h_{ij}) x_j)}{2\sigma_i}  f'(h_{ij}) x_j,
\end{align}
where $f'(h) = \partial f(h)/ \partial h$ is the derivative of the passing probability function.
Combining this result into equation (10) gives the update rule
\begin{align}
    	\Delta h_{ij}  = &- \eta \frac{\partial E(\tilde{y})}{\partial \tilde{y}_i} f'(h_{ij}) x_j \nonumber \\
                       & \times \left( 1 + \frac{1- 2f(h_{ij}) x_j}{2\sigma_i}  \xi_i \right). 
\end{align}
In this form the rule contains the mean field component multiplied by a factor that depends on the variance. While for the derivation of the rule we have specified that $\xi_i$ is a Guassian random variable with zero mean and unit variance, during training it is calculated exactly from the forward phase using $\xi_i = (y_i - \mu_i) / \sigma_i$, so if the neuron output is higher than the mean it will be positive while if it is lower it will be negative. This combines with the $1-2f(h_{ij}) x_j$ to determine whether the factor increases the weight update or reduces it.

\subsection*{Model training details}
As a benchmark we use the MNIST dataset \cite{lecun1998mnist} but to reduce the number of synaptic operations for the experimental hardware it was downsampled by using the MaxPool operation with a filter size of 2x2. This created a set of 14 x 14 pixel images which were mapped to a binary input by thresholding the pixel intensity at 0.5.

The training part of the dataset was randomly split into a 50,000 training and 10,000 validation subsets. A real valued bias was applied output of the simulated binary synapses and these values were converted into a probability using the Softmax function with the loss against the image labels measured using Cross-Entropy loss. Training was performed using mini-batches of 50 images, and iterated until the validation loss did not decrease over 20 epochs. The model with the lowest validation error before the end of training was returned as the trained model. The Adam optimiser was used with a learning rate $\eta = 0.001$ for $K \geq 2$ and $\eta = 0.01$ for $K=1$, determined based on the lower validation error.

\subsection*{On device machine learning testing}

For the demonstration of our stochastic network \emph{in materia} we have used an automated control system to inject a domain wall into the magnetic nanowire at the desired magnetic field given by the synaptic weights. We first optimised the synaptic magnetic fields for our network models in simulation for the cases of $K_\mathrm{train}=1$ and 128, using the method detailed below. For each $K_\mathrm{train}$, we trained 5 models before selecting the model that had the lowest error on the validation dataset. We then transferred these to the hardware with the control software loading the pixel binary values ($x_j$) from the test dataset and using the simulation trained magnetic fields ($h_{ij}$) to control the magnetic synapses. As detailed in figure \ref{fig:architecture}, if the pixel value was 1 the control system would determine whether the domain wall has  pinned or passed the defect site and return a 0 or 1 respectively. The result of this synaptic operation was then passed back to the program running the neural network inference, which computed the neuron values to predict the correct class of the test data.

\begin{acknowledgments}
The authors gratefully acknowledge funding from EPSRC (Grant: EP/S009647/1) and Leverhulme Trust (Research Grant: RPG-2019-097). The authors would like to thank Luca Manneschi and Ian Vidamour for their help and feedback on this work.

\end{acknowledgments}

\section*{Code Availability}
Code related to this paper is available at \url{https://github.com/mattoaellis/binary_stochastic_synapses}.

\appendix

\section{Mean and variance of the Poisson-Binomial distribution}\label{secA1}

For a network of binary stochastic synapses, the output of each synapses is assumed to be an independent random binary event (Bernoulli trial). If these had the same probability then the sum of these events would result in the Binomial distribution but as each synapse has a different input and synaptic probability the value of the neuron will follow a Poisson-Binomial distribution. Since this distribution can be complex to calculate in full we approximate the distribution by a Gaussian \cite{hong2013computing}. The mean of the neuron output, $y_i$, for a given number of samples $K$ is
\begin{align}
    \mu_i & = \mathbb{E}[y_i] \\
          & = \mathbb{E}\left[ \frac{1}{K}  \sum_k \sum_j  w_{ij}^{(k)} x_j\right] \\
          & =  \frac{1}{K} \sum_k  \sum_j  \mathbb{E}[w_{ij}^{(k)} x_j] \\
          & = \frac{1}{K} \sum_k \sum_j \left( 0 f(h_{ij}) x_j + 1 f(h_{ij}) x_j \right) \\
          & =  \sum_j f(h_{ij}) x_j.
\end{align}
where in the final step here we note that the synaptic passing probability $f(h_{ij})$ is independent of $k$. The variance of the distribution is 
\begin{align}
    \sigma_i^2 & = \mathrm{var}[y_i] \\
             & = \mathrm{var}\left[ \frac{1}{K}  \sum_k \sum_j  w_{ij}^{(k)} x_j\right].
\end{align} 
We now use the fact that the variance of a sum of independent random events is the sum of the variances, and that $\mathrm{var}\left[y/K\right] = \mathrm{var}[y]/K^2$ such that
\begin{align}
    \sigma_i^2 & = \frac{1}{K^2} \sum_k \sum_j \mathrm{var}\left[ w_{ij}^{(k)} x_j\right].
\end{align}
The variance of each synapse as a Bernoulli event is
\begin{align}
    \mathrm{var}\left[ w_{ij}^{(k)} x_j\right] & = f(h_{ij}) x_j \left( 1 - f(h_{ij}) x_j \right),
\end{align}
and again since this is independent of $k$ then the variance of the neuron output is
\begin{align}
    \sigma_i^2 & = \frac{K}{K^2} \sum_j f(h_{ij}) x_j \left( 1 - f(h_{ij}) x_j \right) \\
    & = \frac{1}{K} \sum_j f(h_{ij}) x_j \left( 1 - f(h_{ij}) x_j \right).
\end{align}

\draftinfo{
\section*{\hl{Some useful references}}
\textit{\hl{Obviously this section is temporary!}}
MQCA nanomagnetic logic very promising: `non-volatile, dense, low-power, and radiation-hard'~\cite{niemierNanomagnetLogicProgress2011}.
Galton board paper...previous work has demonstrated the predetermined stochastic distributions...~\cite{sanz-hernandezTunableStochasticityArtificial2021}.
Low energy BNNs~\cite{laborieuxLowPowerInMemory2020,simonsReviewBinarizedNeural2019}.
Damien's first paper (building on BNNs, but binarising input as stochastic bit streams and examining it with a hardware design with STT ram)~\cite{hirtzlin2019stochastic}, the second (which we might use as an example of how hardware such as ours could fit into an ASIC design)~\cite{penkovskyInMemoryResistiveRAM2020}.
Damien's papers using DW position in NW to code multiple state quantized weights for NNs~\cite{azam2020voltage,misbaEnergyEfficientLearning2022}.
Daniels' paper, combining fixed value random bit streams digitally to generate random numbers for stochastic neural networks - stochastic neural networks, but not stochastic synapses~\cite{daniels_energy-efficient_2020}. NB: Encoding our inputs as stochastic streams might give us the advantages of both paradigms. 
More papers that address the success of stochastic (input) NNs \cite{liHEIFHighlyEfficient2019,shao2021implementation}, all CMOS implemntation with resulting difficulties, but demonstrating it can be done low-power and compactly~\cite{muthappaHardwarebasedFastRealtime2020}.
Stochastic (input) NN with neuron non-linearity provided by sigmoid from MTJ (effectively just acting as a stochastic generator for the next layer, but making the non-linear calculation easier)~\cite{nisarImplementationEfficientMagnetic2020}.}





\bibliography{refs.bib}

\begin{thebibliography}{64}%
\makeatletter
\providecommand \@ifxundefined [1]{%
 \@ifx{#1\undefined}
}%
\providecommand \@ifnum [1]{%
 \ifnum #1\expandafter \@firstoftwo
 \else \expandafter \@secondoftwo
 \fi
}%
\providecommand \@ifx [1]{%
 \ifx #1\expandafter \@firstoftwo
 \else \expandafter \@secondoftwo
 \fi
}%
\providecommand \natexlab [1]{#1}%
\providecommand \enquote  [1]{``#1''}%
\providecommand \bibnamefont  [1]{#1}%
\providecommand \bibfnamefont [1]{#1}%
\providecommand \citenamefont [1]{#1}%
\providecommand \href@noop [0]{\@secondoftwo}%
\providecommand \href [0]{\begingroup \@sanitize@url \@href}%
\providecommand \@href[1]{\@@startlink{#1}\@@href}%
\providecommand \@@href[1]{\endgroup#1\@@endlink}%
\providecommand \@sanitize@url [0]{\catcode `\\12\catcode `\$12\catcode
  `\&12\catcode `\#12\catcode `\^12\catcode `\_12\catcode `\%12\relax}%
\providecommand \@@startlink[1]{}%
\providecommand \@@endlink[0]{}%
\providecommand \url  [0]{\begingroup\@sanitize@url \@url }%
\providecommand \@url [1]{\endgroup\@href {#1}{\urlprefix }}%
\providecommand \urlprefix  [0]{URL }%
\providecommand \Eprint [0]{\href }%
\providecommand \doibase [0]{http://dx.doi.org/}%
\providecommand \selectlanguage [0]{\@gobble}%
\providecommand \bibinfo  [0]{\@secondoftwo}%
\providecommand \bibfield  [0]{\@secondoftwo}%
\providecommand \translation [1]{[#1]}%
\providecommand \BibitemOpen [0]{}%
\providecommand \bibitemStop [0]{}%
\providecommand \bibitemNoStop [0]{.\EOS\space}%
\providecommand \EOS [0]{\spacefactor3000\relax}%
\providecommand \BibitemShut  [1]{\csname bibitem#1\endcsname}%
\let\auto@bib@innerbib\@empty
\bibitem [{\citenamefont {Thompson}\ \emph {et~al.}(2021)\citenamefont
  {Thompson}, \citenamefont {Greenewald}, \citenamefont {Lee},\ and\
  \citenamefont {Manso}}]{Thompson2021}%
  \BibitemOpen
  \bibfield  {author} {\bibinfo {author} {\bibfnamefont {N.~C.}\ \bibnamefont
  {Thompson}}, \bibinfo {author} {\bibfnamefont {K.}~\bibnamefont
  {Greenewald}}, \bibinfo {author} {\bibfnamefont {K.}~\bibnamefont {Lee}}, \
  and\ \bibinfo {author} {\bibfnamefont {G.~F.}\ \bibnamefont {Manso}},\
  }\bibfield  {title} {\enquote {\bibinfo {title} {Deep learning's diminishing
  returns: The cost of improvement is becoming unsustainable},}\ }\href@noop {}
  {\bibfield  {journal} {\bibinfo  {journal} {IEEE Spectrum}\ }\textbf
  {\bibinfo {volume} {58}},\ \bibinfo {pages} {50--55} (\bibinfo {year}
  {2021})}\BibitemShut {NoStop}%
\bibitem [{\citenamefont {{Sabry Aly}}\ \emph {et~al.}(2019)\citenamefont
  {{Sabry Aly}}, \citenamefont {{Wu}}, \citenamefont {{Bartolo}}, \citenamefont
  {{Malviya}}, \citenamefont {{Hwang}}, \citenamefont {{Hills}}, \citenamefont
  {{Markov}}, \citenamefont {{Wootters}}, \citenamefont {{Shulaker}},
  \citenamefont {{Philip Wong}},\ and\ \citenamefont
  {{Mitra}}}]{SabryAly2019N3XT}%
  \BibitemOpen
  \bibfield  {author} {\bibinfo {author} {\bibfnamefont {M.~M.}\ \bibnamefont
  {{Sabry Aly}}}, \bibinfo {author} {\bibfnamefont {T.~F.}\ \bibnamefont
  {{Wu}}}, \bibinfo {author} {\bibfnamefont {A.}~\bibnamefont {{Bartolo}}},
  \bibinfo {author} {\bibfnamefont {Y.~H.}\ \bibnamefont {{Malviya}}}, \bibinfo
  {author} {\bibfnamefont {W.}~\bibnamefont {{Hwang}}}, \bibinfo {author}
  {\bibfnamefont {G.}~\bibnamefont {{Hills}}}, \bibinfo {author} {\bibfnamefont
  {I.}~\bibnamefont {{Markov}}}, \bibinfo {author} {\bibfnamefont
  {M.}~\bibnamefont {{Wootters}}}, \bibinfo {author} {\bibfnamefont {M.~M.}\
  \bibnamefont {{Shulaker}}}, \bibinfo {author} {\bibfnamefont
  {H.}~\bibnamefont {{Philip Wong}}}, \ and\ \bibinfo {author} {\bibfnamefont
  {S.}~\bibnamefont {{Mitra}}},\ }\bibfield  {title} {\enquote {\bibinfo
  {title} {The n3xt approach to energy-efficient abundant-data computing},}\
  }\href@noop {} {\bibfield  {journal} {\bibinfo  {journal} {Proceedings of the
  IEEE}\ }\textbf {\bibinfo {volume} {107}},\ \bibinfo {pages} {19--48}
  (\bibinfo {year} {2019})}\BibitemShut {NoStop}%
\bibitem [{\citenamefont {Strubell}, \citenamefont {Ganesh},\ and\
  \citenamefont {McCallum}(2019)}]{Strubell2019_NLPenergy}%
  \BibitemOpen
  \bibfield  {author} {\bibinfo {author} {\bibfnamefont {E.}~\bibnamefont
  {Strubell}}, \bibinfo {author} {\bibfnamefont {A.}~\bibnamefont {Ganesh}}, \
  and\ \bibinfo {author} {\bibfnamefont {A.}~\bibnamefont {McCallum}},\ }\href
  {https://arxiv.org/abs/1906.02243} {\enquote {\bibinfo {title} {Energy and
  policy considerations for deep learning in nlp},}\ } (\bibinfo {year}
  {2019})\BibitemShut {NoStop}%
\bibitem [{\citenamefont {Ziegler}(2020)}]{ziegler2020novel}%
  \BibitemOpen
  \bibfield  {author} {\bibinfo {author} {\bibfnamefont {M.}~\bibnamefont
  {Ziegler}},\ }\bibfield  {title} {\enquote {\bibinfo {title} {Novel hardware
  and concepts for unconventional computing},}\ }\href@noop {} {\bibfield
  {journal} {\bibinfo  {journal} {Scientific reports}\ }\textbf {\bibinfo
  {volume} {10}},\ \bibinfo {pages} {1--3} (\bibinfo {year}
  {2020})}\BibitemShut {NoStop}%
\bibitem [{\citenamefont {Niemier}\ \emph {et~al.}(2011)\citenamefont
  {Niemier}, \citenamefont {Bernstein}, \citenamefont {Csaba}, \citenamefont
  {Dingler}, \citenamefont {Hu}, \citenamefont {Kurtz}, \citenamefont {Liu},
  \citenamefont {Nahas}, \citenamefont {Porod}, \citenamefont {Siddiq},\ and\
  \citenamefont {Varga}}]{niemierNanomagnetLogicProgress2011}%
  \BibitemOpen
  \bibfield  {author} {\bibinfo {author} {\bibfnamefont {M.~T.}\ \bibnamefont
  {Niemier}}, \bibinfo {author} {\bibfnamefont {G.~H.}\ \bibnamefont
  {Bernstein}}, \bibinfo {author} {\bibfnamefont {G.}~\bibnamefont {Csaba}},
  \bibinfo {author} {\bibfnamefont {A.}~\bibnamefont {Dingler}}, \bibinfo
  {author} {\bibfnamefont {X.~S.}\ \bibnamefont {Hu}}, \bibinfo {author}
  {\bibfnamefont {S.}~\bibnamefont {Kurtz}}, \bibinfo {author} {\bibfnamefont
  {S.}~\bibnamefont {Liu}}, \bibinfo {author} {\bibfnamefont {J.}~\bibnamefont
  {Nahas}}, \bibinfo {author} {\bibfnamefont {W.}~\bibnamefont {Porod}},
  \bibinfo {author} {\bibfnamefont {M.}~\bibnamefont {Siddiq}}, \ and\ \bibinfo
  {author} {\bibfnamefont {E.}~\bibnamefont {Varga}},\ }\bibfield  {title}
  {\enquote {\bibinfo {title} {Nanomagnet logic: Progress toward system-level
  integration},}\ }\href {https://dx.doi.org/10.1088/0953-8984/23/49/493202}
  {\bibfield  {journal} {\bibinfo  {journal} {Journal of Physics: Condensed
  Matter}\ }\textbf {\bibinfo {volume} {23}},\ \bibinfo {pages} {493202}
  (\bibinfo {year} {2011})}\BibitemShut {NoStop}%
\bibitem [{\citenamefont {Finocchio}\ \emph {et~al.}(2021)\citenamefont
  {Finocchio}, \citenamefont {Di~Ventra}, \citenamefont {Camsari},
  \citenamefont {Everschor-Sitte}, \citenamefont {Amiri},\ and\ \citenamefont
  {Zeng}}]{finocchio2021promise}%
  \BibitemOpen
  \bibfield  {author} {\bibinfo {author} {\bibfnamefont {G.}~\bibnamefont
  {Finocchio}}, \bibinfo {author} {\bibfnamefont {M.}~\bibnamefont
  {Di~Ventra}}, \bibinfo {author} {\bibfnamefont {K.~Y.}\ \bibnamefont
  {Camsari}}, \bibinfo {author} {\bibfnamefont {K.}~\bibnamefont
  {Everschor-Sitte}}, \bibinfo {author} {\bibfnamefont {P.~K.}\ \bibnamefont
  {Amiri}}, \ and\ \bibinfo {author} {\bibfnamefont {Z.}~\bibnamefont {Zeng}},\
  }\bibfield  {title} {\enquote {\bibinfo {title} {The promise of spintronics
  for unconventional computing},}\ }\href@noop {} {\bibfield  {journal}
  {\bibinfo  {journal} {Journal of Magnetism and Magnetic Materials}\ }\textbf
  {\bibinfo {volume} {521}},\ \bibinfo {pages} {167506} (\bibinfo {year}
  {2021})}\BibitemShut {NoStop}%
\bibitem [{\citenamefont {Grollier}\ \emph {et~al.}(2020)\citenamefont
  {Grollier}, \citenamefont {Querlioz}, \citenamefont {Camsari}, \citenamefont
  {Everschor-Sitte}, \citenamefont {Fukami},\ and\ \citenamefont
  {Stiles}}]{Grollier2020NeuroSpintron}%
  \BibitemOpen
  \bibfield  {author} {\bibinfo {author} {\bibfnamefont {J.}~\bibnamefont
  {Grollier}}, \bibinfo {author} {\bibfnamefont {D.}~\bibnamefont {Querlioz}},
  \bibinfo {author} {\bibfnamefont {K.~Y.}\ \bibnamefont {Camsari}}, \bibinfo
  {author} {\bibfnamefont {K.}~\bibnamefont {Everschor-Sitte}}, \bibinfo
  {author} {\bibfnamefont {S.}~\bibnamefont {Fukami}}, \ and\ \bibinfo {author}
  {\bibfnamefont {M.~D.}\ \bibnamefont {Stiles}},\ }\bibfield  {title}
  {\enquote {\bibinfo {title} {Neuromorphic spintronics},}\ }\href
  {https://doi.org/10.1038/s41928-019-0360-9} {\bibfield  {journal} {\bibinfo
  {journal} {Nature Electronics}\ }\textbf {\bibinfo {volume} {3}},\ \bibinfo
  {pages} {360--370} (\bibinfo {year} {2020})}\BibitemShut {NoStop}%
\bibitem [{\citenamefont {Petersen}\ \emph {et~al.}(1998)\citenamefont
  {Petersen}, \citenamefont {Malenka}, \citenamefont {Nicoll},\ and\
  \citenamefont {Hopfield}}]{petersen1998all}%
  \BibitemOpen
  \bibfield  {author} {\bibinfo {author} {\bibfnamefont {C.~C.}\ \bibnamefont
  {Petersen}}, \bibinfo {author} {\bibfnamefont {R.~C.}\ \bibnamefont
  {Malenka}}, \bibinfo {author} {\bibfnamefont {R.~A.}\ \bibnamefont {Nicoll}},
  \ and\ \bibinfo {author} {\bibfnamefont {J.~J.}\ \bibnamefont {Hopfield}},\
  }\bibfield  {title} {\enquote {\bibinfo {title} {All-or-none potentiation at
  ca3-ca1 synapses},}\ }\href@noop {} {\bibfield  {journal} {\bibinfo
  {journal} {Proceedings of the National Academy of Sciences}\ }\textbf
  {\bibinfo {volume} {95}},\ \bibinfo {pages} {4732--4737} (\bibinfo {year}
  {1998})}\BibitemShut {NoStop}%
\bibitem [{\citenamefont {Simons}\ and\ \citenamefont
  {Lee}(2019)}]{simonsReviewBinarizedNeural2019}%
  \BibitemOpen
  \bibfield  {author} {\bibinfo {author} {\bibfnamefont {T.}~\bibnamefont
  {Simons}}\ and\ \bibinfo {author} {\bibfnamefont {D.-J.}\ \bibnamefont
  {Lee}},\ }\bibfield  {title} {\enquote {\bibinfo {title} {A {{Review}} of
  {{Binarized Neural Networks}}},}\ }\href
  {https://www.mdpi.com/2079-9292/8/6/661} {\bibfield  {journal} {\bibinfo
  {journal} {Electronics}\ }\textbf {\bibinfo {volume} {8}},\ \bibinfo {pages}
  {661} (\bibinfo {year} {2019})}\BibitemShut {NoStop}%
\bibitem [{\citenamefont {Laborieux}\ \emph {et~al.}(2020)\citenamefont
  {Laborieux}, \citenamefont {Bocquet}, \citenamefont {Hirtzlin}, \citenamefont
  {Klein}, \citenamefont {Diez}, \citenamefont {Nowak}, \citenamefont
  {Vianello}, \citenamefont {Portal},\ and\ \citenamefont
  {Querlioz}}]{laborieuxLowPowerInMemory2020}%
  \BibitemOpen
  \bibfield  {author} {\bibinfo {author} {\bibfnamefont {A.}~\bibnamefont
  {Laborieux}}, \bibinfo {author} {\bibfnamefont {M.}~\bibnamefont {Bocquet}},
  \bibinfo {author} {\bibfnamefont {T.}~\bibnamefont {Hirtzlin}}, \bibinfo
  {author} {\bibfnamefont {J.-O.}\ \bibnamefont {Klein}}, \bibinfo {author}
  {\bibfnamefont {L.~H.}\ \bibnamefont {Diez}}, \bibinfo {author}
  {\bibfnamefont {E.}~\bibnamefont {Nowak}}, \bibinfo {author} {\bibfnamefont
  {E.}~\bibnamefont {Vianello}}, \bibinfo {author} {\bibfnamefont {J.-M.}\
  \bibnamefont {Portal}}, \ and\ \bibinfo {author} {\bibfnamefont
  {D.}~\bibnamefont {Querlioz}},\ }\bibfield  {title} {\enquote {\bibinfo
  {title} {Low {{Power In-Memory Implementation}} of {{Ternary Neural
  Networks}} with {{Resistive RAM-Based Synapse}}},}\ }in\ \href@noop {} {\emph
  {\bibinfo {booktitle} {2020 2nd {{IEEE International Conference}} on
  {{Artificial Intelligence Circuits}} and {{Systems}} ({{AICAS}})}}}\
  (\bibinfo {year} {2020})\ pp.\ \bibinfo {pages} {136--140}\BibitemShut
  {NoStop}%
\bibitem [{\citenamefont {Yu}\ \emph {et~al.}(2013)\citenamefont {Yu},
  \citenamefont {Gao}, \citenamefont {Fang}, \citenamefont {Yu}, \citenamefont
  {Kang},\ and\ \citenamefont {Wong}}]{yu2013stochastic}%
  \BibitemOpen
  \bibfield  {author} {\bibinfo {author} {\bibfnamefont {S.}~\bibnamefont
  {Yu}}, \bibinfo {author} {\bibfnamefont {B.}~\bibnamefont {Gao}}, \bibinfo
  {author} {\bibfnamefont {Z.}~\bibnamefont {Fang}}, \bibinfo {author}
  {\bibfnamefont {H.}~\bibnamefont {Yu}}, \bibinfo {author} {\bibfnamefont
  {J.}~\bibnamefont {Kang}}, \ and\ \bibinfo {author} {\bibfnamefont
  {H.-S.~P.}\ \bibnamefont {Wong}},\ }\bibfield  {title} {\enquote {\bibinfo
  {title} {Stochastic learning in oxide binary synaptic device for neuromorphic
  computing},}\ }\href@noop {} {\bibfield  {journal} {\bibinfo  {journal}
  {Frontiers in neuroscience}\ }\textbf {\bibinfo {volume} {7}},\ \bibinfo
  {pages} {186} (\bibinfo {year} {2013})}\BibitemShut {NoStop}%
\bibitem [{\citenamefont {Penkovsky}\ \emph {et~al.}(2020)\citenamefont
  {Penkovsky}, \citenamefont {Bocquet}, \citenamefont {Hirtzlin}, \citenamefont
  {Klein}, \citenamefont {Nowak}, \citenamefont {Vianello}, \citenamefont
  {Portal},\ and\ \citenamefont
  {Querlioz}}]{penkovskyInMemoryResistiveRAM2020}%
  \BibitemOpen
  \bibfield  {author} {\bibinfo {author} {\bibfnamefont {B.}~\bibnamefont
  {Penkovsky}}, \bibinfo {author} {\bibfnamefont {M.}~\bibnamefont {Bocquet}},
  \bibinfo {author} {\bibfnamefont {T.}~\bibnamefont {Hirtzlin}}, \bibinfo
  {author} {\bibfnamefont {J.-O.}\ \bibnamefont {Klein}}, \bibinfo {author}
  {\bibfnamefont {E.}~\bibnamefont {Nowak}}, \bibinfo {author} {\bibfnamefont
  {E.}~\bibnamefont {Vianello}}, \bibinfo {author} {\bibfnamefont {J.-M.}\
  \bibnamefont {Portal}}, \ and\ \bibinfo {author} {\bibfnamefont
  {D.}~\bibnamefont {Querlioz}},\ }\bibfield  {title} {\enquote {\bibinfo
  {title} {In-{{Memory Resistive RAM Implementation}} of {{Binarized Neural
  Networks}} for {{Medical Applications}}},}\ }in\ \href@noop {} {\emph
  {\bibinfo {booktitle} {2020 {{Design}}, {{Automation}} \& {{Test}} in
  {{Europe Conference}} \& {{Exhibition}}}}}\ (\bibinfo {year} {2020})\ pp.\
  \bibinfo {pages} {690--695}\BibitemShut {NoStop}%
\bibitem [{\citenamefont {Neftci}\ \emph {et~al.}(2016)\citenamefont {Neftci},
  \citenamefont {Pedroni}, \citenamefont {Joshi}, \citenamefont {Al-Shedivat},\
  and\ \citenamefont {Cauwenberghs}}]{neftci2016stochastic}%
  \BibitemOpen
  \bibfield  {author} {\bibinfo {author} {\bibfnamefont {E.~O.}\ \bibnamefont
  {Neftci}}, \bibinfo {author} {\bibfnamefont {B.~U.}\ \bibnamefont {Pedroni}},
  \bibinfo {author} {\bibfnamefont {S.}~\bibnamefont {Joshi}}, \bibinfo
  {author} {\bibfnamefont {M.}~\bibnamefont {Al-Shedivat}}, \ and\ \bibinfo
  {author} {\bibfnamefont {G.}~\bibnamefont {Cauwenberghs}},\ }\bibfield
  {title} {\enquote {\bibinfo {title} {Stochastic synapses enable efficient
  brain-inspired learning machines},}\ }\href@noop {} {\bibfield  {journal}
  {\bibinfo  {journal} {Frontiers in neuroscience}\ }\textbf {\bibinfo {volume}
  {10}},\ \bibinfo {pages} {241} (\bibinfo {year} {2016})}\BibitemShut
  {NoStop}%
\bibitem [{\citenamefont {Daniels}\ \emph {et~al.}(2020)\citenamefont
  {Daniels}, \citenamefont {Madhavan}, \citenamefont {Talatchian},
  \citenamefont {Mizrahi},\ and\ \citenamefont
  {Stiles}}]{daniels_energy-efficient_2020}%
  \BibitemOpen
  \bibfield  {author} {\bibinfo {author} {\bibfnamefont {M.~W.}\ \bibnamefont
  {Daniels}}, \bibinfo {author} {\bibfnamefont {A.}~\bibnamefont {Madhavan}},
  \bibinfo {author} {\bibfnamefont {P.}~\bibnamefont {Talatchian}}, \bibinfo
  {author} {\bibfnamefont {A.}~\bibnamefont {Mizrahi}}, \ and\ \bibinfo
  {author} {\bibfnamefont {M.~D.}\ \bibnamefont {Stiles}},\ }\bibfield  {title}
  {\enquote {\bibinfo {title} {Energy-efficient stochastic computing with
  superparamagnetic tunnel junctions},}\ }\href {\doibase
  10.1103/PhysRevApplied.13.034016} {\bibfield  {journal} {\bibinfo  {journal}
  {Phys. Rev. Appl.}\ }\textbf {\bibinfo {volume} {13}},\ \bibinfo {pages}
  {034016} (\bibinfo {year} {2020})}\BibitemShut {NoStop}%
\bibitem [{\citenamefont {Li}\ \emph {et~al.}(2019)\citenamefont {Li},
  \citenamefont {Li}, \citenamefont {Ren}, \citenamefont {Cai}, \citenamefont
  {Ding}, \citenamefont {Qian}, \citenamefont {Draper}, \citenamefont {Yuan},
  \citenamefont {Tang}, \citenamefont {Qiu},\ and\ \citenamefont
  {Wang}}]{liHEIFHighlyEfficient2019}%
  \BibitemOpen
  \bibfield  {author} {\bibinfo {author} {\bibfnamefont {Z.}~\bibnamefont
  {Li}}, \bibinfo {author} {\bibfnamefont {J.}~\bibnamefont {Li}}, \bibinfo
  {author} {\bibfnamefont {A.}~\bibnamefont {Ren}}, \bibinfo {author}
  {\bibfnamefont {R.}~\bibnamefont {Cai}}, \bibinfo {author} {\bibfnamefont
  {C.}~\bibnamefont {Ding}}, \bibinfo {author} {\bibfnamefont {X.}~\bibnamefont
  {Qian}}, \bibinfo {author} {\bibfnamefont {J.}~\bibnamefont {Draper}},
  \bibinfo {author} {\bibfnamefont {B.}~\bibnamefont {Yuan}}, \bibinfo {author}
  {\bibfnamefont {J.}~\bibnamefont {Tang}}, \bibinfo {author} {\bibfnamefont
  {Q.}~\bibnamefont {Qiu}}, \ and\ \bibinfo {author} {\bibfnamefont
  {Y.}~\bibnamefont {Wang}},\ }\bibfield  {title} {\enquote {\bibinfo {title}
  {{{HEIF}}: {{Highly Efficient Stochastic Computing-Based Inference
  Framework}} for {{Deep Neural Networks}}},}\ }\href@noop {} {\bibfield
  {journal} {\bibinfo  {journal} {IEEE Transactions on Computer-Aided Design of
  Integrated Circuits and Systems}\ }\textbf {\bibinfo {volume} {38}},\
  \bibinfo {pages} {1543--1556} (\bibinfo {year} {2019})}\BibitemShut {NoStop}%
\bibitem [{\citenamefont {Shao}\ \emph {et~al.}(2021)\citenamefont {Shao},
  \citenamefont {Sinaga}, \citenamefont {Sunmola}, \citenamefont {Borland},
  \citenamefont {Carey}, \citenamefont {Katine}, \citenamefont
  {Lopez-Dominguez},\ and\ \citenamefont {Amiri}}]{shao2021implementation}%
  \BibitemOpen
  \bibfield  {author} {\bibinfo {author} {\bibfnamefont {Y.}~\bibnamefont
  {Shao}}, \bibinfo {author} {\bibfnamefont {S.~L.}\ \bibnamefont {Sinaga}},
  \bibinfo {author} {\bibfnamefont {I.~O.}\ \bibnamefont {Sunmola}}, \bibinfo
  {author} {\bibfnamefont {A.~S.}\ \bibnamefont {Borland}}, \bibinfo {author}
  {\bibfnamefont {M.~J.}\ \bibnamefont {Carey}}, \bibinfo {author}
  {\bibfnamefont {J.~A.}\ \bibnamefont {Katine}}, \bibinfo {author}
  {\bibfnamefont {V.}~\bibnamefont {Lopez-Dominguez}}, \ and\ \bibinfo {author}
  {\bibfnamefont {P.~K.}\ \bibnamefont {Amiri}},\ }\bibfield  {title} {\enquote
  {\bibinfo {title} {Implementation of artificial neural networks using
  magnetoresistive random-access memory-based stochastic computing units},}\
  }\href@noop {} {\bibfield  {journal} {\bibinfo  {journal} {IEEE Magnetics
  Letters}\ }\textbf {\bibinfo {volume} {12}},\ \bibinfo {pages} {1--5}
  (\bibinfo {year} {2021})}\BibitemShut {NoStop}%
\bibitem [{\citenamefont {Muthappa}\ \emph {et~al.}(2020)\citenamefont
  {Muthappa}, \citenamefont {Neugebauer}, \citenamefont {Polian},\ and\
  \citenamefont {Hayes}}]{muthappaHardwarebasedFastRealtime2020}%
  \BibitemOpen
  \bibfield  {author} {\bibinfo {author} {\bibfnamefont {P.~K.}\ \bibnamefont
  {Muthappa}}, \bibinfo {author} {\bibfnamefont {F.}~\bibnamefont
  {Neugebauer}}, \bibinfo {author} {\bibfnamefont {I.}~\bibnamefont {Polian}},
  \ and\ \bibinfo {author} {\bibfnamefont {J.~P.}\ \bibnamefont {Hayes}},\
  }\bibfield  {title} {\enquote {\bibinfo {title} {Hardware-based {{Fast
  Real-time Image Classification}} with {{Stochastic Computing}}},}\ }in\
  \href@noop {} {\emph {\bibinfo {booktitle} {2020 {{IEEE}} 38th
  {{International Conference}} on {{Computer Design}} ({{ICCD}})}}}\ (\bibinfo
  {year} {2020})\ pp.\ \bibinfo {pages} {340--347}\BibitemShut {NoStop}%
\bibitem [{\citenamefont {Nisar}, \citenamefont {Khanday},\ and\ \citenamefont
  {Kaushik}(2020)}]{nisarImplementationEfficientMagnetic2020}%
  \BibitemOpen
  \bibfield  {author} {\bibinfo {author} {\bibfnamefont {A.}~\bibnamefont
  {Nisar}}, \bibinfo {author} {\bibfnamefont {F.~A.}\ \bibnamefont {Khanday}},
  \ and\ \bibinfo {author} {\bibfnamefont {B.~K.}\ \bibnamefont {Kaushik}},\
  }\bibfield  {title} {\enquote {\bibinfo {title} {Implementation of an
  efficient magnetic tunnel junction-based stochastic neural network with
  application to iris data classification},}\ }\href
  {https://dx.doi.org/10.1088/1361-6528/abadc4} {\bibfield  {journal} {\bibinfo
   {journal} {Nanotechnology}\ }\textbf {\bibinfo {volume} {31}},\ \bibinfo
  {pages} {504001} (\bibinfo {year} {2020})}\BibitemShut {NoStop}%
\bibitem [{\citenamefont {Hirtzlin}\ \emph {et~al.}(2019)\citenamefont
  {Hirtzlin}, \citenamefont {Penkovsky}, \citenamefont {Bocquet}, \citenamefont
  {Klein}, \citenamefont {Portal},\ and\ \citenamefont
  {Querlioz}}]{hirtzlin2019stochastic}%
  \BibitemOpen
  \bibfield  {author} {\bibinfo {author} {\bibfnamefont {T.}~\bibnamefont
  {Hirtzlin}}, \bibinfo {author} {\bibfnamefont {B.}~\bibnamefont {Penkovsky}},
  \bibinfo {author} {\bibfnamefont {M.}~\bibnamefont {Bocquet}}, \bibinfo
  {author} {\bibfnamefont {J.-O.}\ \bibnamefont {Klein}}, \bibinfo {author}
  {\bibfnamefont {J.-M.}\ \bibnamefont {Portal}}, \ and\ \bibinfo {author}
  {\bibfnamefont {D.}~\bibnamefont {Querlioz}},\ }\bibfield  {title} {\enquote
  {\bibinfo {title} {Stochastic computing for hardware implementation of
  binarized neural networks},}\ }\href@noop {} {\bibfield  {journal} {\bibinfo
  {journal} {IEEE Access}\ }\textbf {\bibinfo {volume} {7}},\ \bibinfo {pages}
  {76394--76403} (\bibinfo {year} {2019})}\BibitemShut {NoStop}%
\bibitem [{\citenamefont {Baibich}\ \emph {et~al.}(1988)\citenamefont
  {Baibich}, \citenamefont {Broto}, \citenamefont {Fert}, \citenamefont
  {Van~Dau}, \citenamefont {Petroff}, \citenamefont {Etienne}, \citenamefont
  {Creuzet}, \citenamefont {Friederich},\ and\ \citenamefont
  {Chazelas}}]{Baibich1988}%
  \BibitemOpen
  \bibfield  {author} {\bibinfo {author} {\bibfnamefont {M.~N.}\ \bibnamefont
  {Baibich}}, \bibinfo {author} {\bibfnamefont {J.~M.}\ \bibnamefont {Broto}},
  \bibinfo {author} {\bibfnamefont {A.}~\bibnamefont {Fert}}, \bibinfo {author}
  {\bibfnamefont {F.~N.}\ \bibnamefont {Van~Dau}}, \bibinfo {author}
  {\bibfnamefont {F.}~\bibnamefont {Petroff}}, \bibinfo {author} {\bibfnamefont
  {P.}~\bibnamefont {Etienne}}, \bibinfo {author} {\bibfnamefont
  {G.}~\bibnamefont {Creuzet}}, \bibinfo {author} {\bibfnamefont
  {A.}~\bibnamefont {Friederich}}, \ and\ \bibinfo {author} {\bibfnamefont
  {J.}~\bibnamefont {Chazelas}},\ }\bibfield  {title} {\enquote {\bibinfo
  {title} {Giant {Magnetoresistance} of (001){Fe}/(001){Cr} {Magnetic}
  {Superlattices}},}\ }\href
  {https://link.aps.org/doi/10.1103/PhysRevLett.61.2472} {\bibfield  {journal}
  {\bibinfo  {journal} {Physical Review Letters}\ }\textbf {\bibinfo {volume}
  {61}},\ \bibinfo {pages} {2472--2475} (\bibinfo {year} {1988})}\BibitemShut
  {NoStop}%
\bibitem [{\citenamefont {Binasch}\ \emph {et~al.}(1989)\citenamefont
  {Binasch}, \citenamefont {Grünberg}, \citenamefont {Saurenbach},\ and\
  \citenamefont {Zinn}}]{binaschEnhancedMagnetoresistanceLayered1989}%
  \BibitemOpen
  \bibfield  {author} {\bibinfo {author} {\bibfnamefont {G.}~\bibnamefont
  {Binasch}}, \bibinfo {author} {\bibfnamefont {P.}~\bibnamefont {Grünberg}},
  \bibinfo {author} {\bibfnamefont {F.}~\bibnamefont {Saurenbach}}, \ and\
  \bibinfo {author} {\bibfnamefont {W.}~\bibnamefont {Zinn}},\ }\bibfield
  {title} {\enquote {\bibinfo {title} {Enhanced magnetoresistance in layered
  magnetic structures with antiferromagnetic interlayer exchange},}\ }\href
  {https://link.aps.org/doi/10.1103/PhysRevB.39.4828} {\bibfield  {journal}
  {\bibinfo  {journal} {Physical Review B}\ }\textbf {\bibinfo {volume} {39}},\
  \bibinfo {pages} {4828--4830} (\bibinfo {year} {1989})}\BibitemShut {NoStop}%
\bibitem [{\citenamefont {Allwood}\ \emph {et~al.}(2005)\citenamefont
  {Allwood}, \citenamefont {Xiong}, \citenamefont {Faulkner}, \citenamefont
  {Atkinson}, \citenamefont {Petit},\ and\ \citenamefont
  {Cowburn}}]{allwood2005magnetic}%
  \BibitemOpen
  \bibfield  {author} {\bibinfo {author} {\bibfnamefont {D.~A.}\ \bibnamefont
  {Allwood}}, \bibinfo {author} {\bibfnamefont {G.}~\bibnamefont {Xiong}},
  \bibinfo {author} {\bibfnamefont {C.}~\bibnamefont {Faulkner}}, \bibinfo
  {author} {\bibfnamefont {D.}~\bibnamefont {Atkinson}}, \bibinfo {author}
  {\bibfnamefont {D.}~\bibnamefont {Petit}}, \ and\ \bibinfo {author}
  {\bibfnamefont {R.}~\bibnamefont {Cowburn}},\ }\bibfield  {title} {\enquote
  {\bibinfo {title} {Magnetic domain-wall logic},}\ }\href@noop {} {\bibfield
  {journal} {\bibinfo  {journal} {Science}\ }\textbf {\bibinfo {volume}
  {309}},\ \bibinfo {pages} {1688--1692} (\bibinfo {year} {2005})}\BibitemShut
  {NoStop}%
\bibitem [{\citenamefont {Mccray}(2009)}]{Mccray2009}%
  \BibitemOpen
  \bibfield  {author} {\bibinfo {author} {\bibfnamefont {W.~P.}\ \bibnamefont
  {Mccray}},\ }\bibfield  {title} {\enquote {\bibinfo {title} {How spintronics
  went from the lab to the {iPod}},}\ }\href@noop {} {\bibfield  {journal}
  {\bibinfo  {journal} {Nature Nanotechnology}\ }\textbf {\bibinfo {volume}
  {4}} (\bibinfo {year} {2009})}\BibitemShut {NoStop}%
\bibitem [{\citenamefont {Parkin}, \citenamefont {Hayashi},\ and\ \citenamefont
  {Thomas}(2008{\natexlab{a}})}]{parkinMagneticDomainwallRacetrack2008}%
  \BibitemOpen
  \bibfield  {author} {\bibinfo {author} {\bibfnamefont {S.~S.}\ \bibnamefont
  {Parkin}}, \bibinfo {author} {\bibfnamefont {M.}~\bibnamefont {Hayashi}}, \
  and\ \bibinfo {author} {\bibfnamefont {L.}~\bibnamefont {Thomas}},\
  }\bibfield  {title} {\enquote {\bibinfo {title} {Magnetic domain-wall
  racetrack memory},}\ }\href@noop {} {\bibfield  {journal} {\bibinfo
  {journal} {Science}\ }\textbf {\bibinfo {volume} {320}},\ \bibinfo {pages}
  {190--194} (\bibinfo {year} {2008}{\natexlab{a}})}\BibitemShut {NoStop}%
\bibitem [{\citenamefont {Lavrijsen}\ \emph {et~al.}(2013)\citenamefont
  {Lavrijsen}, \citenamefont {Lee}, \citenamefont {{Fern{\'a}ndez-Pacheco}},
  \citenamefont {Petit}, \citenamefont {Mansell},\ and\ \citenamefont
  {Cowburn}}]{Lavrijsen2013Magnetic-ratche}%
  \BibitemOpen
  \bibfield  {author} {\bibinfo {author} {\bibfnamefont {R.}~\bibnamefont
  {Lavrijsen}}, \bibinfo {author} {\bibfnamefont {J.-H.}\ \bibnamefont {Lee}},
  \bibinfo {author} {\bibfnamefont {A.}~\bibnamefont
  {{Fern{\'a}ndez-Pacheco}}}, \bibinfo {author} {\bibfnamefont {D.~C. M.~C.}\
  \bibnamefont {Petit}}, \bibinfo {author} {\bibfnamefont {R.}~\bibnamefont
  {Mansell}}, \ and\ \bibinfo {author} {\bibfnamefont {R.~P.}\ \bibnamefont
  {Cowburn}},\ }\bibfield  {title} {\enquote {\bibinfo {title} {Magnetic
  ratchet for three-dimensional spintronic memory and logic},}\ }\href
  {http://www.nature.com/doifinder/10.1038/nature11733} {\bibfield  {journal}
  {\bibinfo  {journal} {Nature}\ }\textbf {\bibinfo {volume} {493}},\ \bibinfo
  {pages} {647--650} (\bibinfo {year} {2013})}\BibitemShut {NoStop}%
\bibitem [{\citenamefont {{Fern{\'a}ndez-Pacheco}}\ \emph
  {et~al.}(2016)\citenamefont {{Fern{\'a}ndez-Pacheco}}, \citenamefont
  {Steinke}, \citenamefont {Mahendru}, \citenamefont {Welbourne}, \citenamefont
  {Mansell}, \citenamefont {Chin}, \citenamefont {Petit}, \citenamefont {Lee},
  \citenamefont {Dalgliesh}, \citenamefont {Langridge},\ and\ \citenamefont
  {Cowburn}}]{Fernandez-Pacheco2016b}%
  \BibitemOpen
  \bibfield  {author} {\bibinfo {author} {\bibfnamefont {A.}~\bibnamefont
  {{Fern{\'a}ndez-Pacheco}}}, \bibinfo {author} {\bibfnamefont {N.-J.}\
  \bibnamefont {Steinke}}, \bibinfo {author} {\bibfnamefont {D.}~\bibnamefont
  {Mahendru}}, \bibinfo {author} {\bibfnamefont {A.}~\bibnamefont {Welbourne}},
  \bibinfo {author} {\bibfnamefont {R.}~\bibnamefont {Mansell}}, \bibinfo
  {author} {\bibfnamefont {S.}~\bibnamefont {Chin}}, \bibinfo {author}
  {\bibfnamefont {D.}~\bibnamefont {Petit}}, \bibinfo {author} {\bibfnamefont
  {J.}~\bibnamefont {Lee}}, \bibinfo {author} {\bibfnamefont {R.}~\bibnamefont
  {Dalgliesh}}, \bibinfo {author} {\bibfnamefont {S.}~\bibnamefont
  {Langridge}}, \ and\ \bibinfo {author} {\bibfnamefont {R.}~\bibnamefont
  {Cowburn}},\ }\bibfield  {title} {\enquote {\bibinfo {title} {Magnetic
  {{State}} of {{Multilayered Synthetic Antiferromagnets}} during {{Soliton
  Nucleation}} and {{Propagation}} for {{Vertical Data Transfer}}},}\ }\href
  {https://onlinelibrary.wiley.com/doi/full/10.1002/admi.201600097} {\bibfield
  {journal} {\bibinfo  {journal} {Advanced Materials Interfaces}\ } (\bibinfo
  {year} {2016})}\BibitemShut {NoStop}%
\bibitem [{\citenamefont {Emori}\ \emph {et~al.}(2013)\citenamefont {Emori},
  \citenamefont {Bauer}, \citenamefont {Ahn}, \citenamefont {Martinez},\ and\
  \citenamefont {Beach}}]{emori2013current}%
  \BibitemOpen
  \bibfield  {author} {\bibinfo {author} {\bibfnamefont {S.}~\bibnamefont
  {Emori}}, \bibinfo {author} {\bibfnamefont {U.}~\bibnamefont {Bauer}},
  \bibinfo {author} {\bibfnamefont {S.-M.}\ \bibnamefont {Ahn}}, \bibinfo
  {author} {\bibfnamefont {E.}~\bibnamefont {Martinez}}, \ and\ \bibinfo
  {author} {\bibfnamefont {G.~S.}\ \bibnamefont {Beach}},\ }\bibfield  {title}
  {\enquote {\bibinfo {title} {Current-driven dynamics of chiral ferromagnetic
  domain walls},}\ }\href@noop {} {\bibfield  {journal} {\bibinfo  {journal}
  {Nature materials}\ }\textbf {\bibinfo {volume} {12}},\ \bibinfo {pages}
  {611--616} (\bibinfo {year} {2013})}\BibitemShut {NoStop}%
\bibitem [{\citenamefont {Hu}\ \emph {et~al.}(2011)\citenamefont {Hu},
  \citenamefont {Li}, \citenamefont {Chen},\ and\ \citenamefont
  {Nan}}]{hu2011high}%
  \BibitemOpen
  \bibfield  {author} {\bibinfo {author} {\bibfnamefont {J.-M.}\ \bibnamefont
  {Hu}}, \bibinfo {author} {\bibfnamefont {Z.}~\bibnamefont {Li}}, \bibinfo
  {author} {\bibfnamefont {L.-Q.}\ \bibnamefont {Chen}}, \ and\ \bibinfo
  {author} {\bibfnamefont {C.-W.}\ \bibnamefont {Nan}},\ }\bibfield  {title}
  {\enquote {\bibinfo {title} {High-density magnetoresistive random access
  memory operating at ultralow voltage at room temperature},}\ }\href@noop {}
  {\bibfield  {journal} {\bibinfo  {journal} {Nature communications}\ }\textbf
  {\bibinfo {volume} {2}},\ \bibinfo {pages} {1--8} (\bibinfo {year}
  {2011})}\BibitemShut {NoStop}%
\bibitem [{\citenamefont {Parkin}, \citenamefont {Hayashi},\ and\ \citenamefont
  {Thomas}(2008{\natexlab{b}})}]{parkin2008magnetic}%
  \BibitemOpen
  \bibfield  {author} {\bibinfo {author} {\bibfnamefont {S.~S.}\ \bibnamefont
  {Parkin}}, \bibinfo {author} {\bibfnamefont {M.}~\bibnamefont {Hayashi}}, \
  and\ \bibinfo {author} {\bibfnamefont {L.}~\bibnamefont {Thomas}},\
  }\bibfield  {title} {\enquote {\bibinfo {title} {Magnetic domain-wall
  racetrack memory},}\ }\href@noop {} {\bibfield  {journal} {\bibinfo
  {journal} {Science}\ }\textbf {\bibinfo {volume} {320}},\ \bibinfo {pages}
  {190--194} (\bibinfo {year} {2008}{\natexlab{b}})}\BibitemShut {NoStop}%
\bibitem [{\citenamefont {{Sanz-Hern{\'a}ndez}}\ \emph
  {et~al.}(2017)\citenamefont {{Sanz-Hern{\'a}ndez}}, \citenamefont {Hamans},
  \citenamefont {Liao}, \citenamefont {Welbourne}, \citenamefont {Lavrijsen},\
  and\ \citenamefont {{Fern{\'a}ndez-Pacheco}}}]{Sanz-Hernandez2017a}%
  \BibitemOpen
  \bibfield  {author} {\bibinfo {author} {\bibfnamefont {D.}~\bibnamefont
  {{Sanz-Hern{\'a}ndez}}}, \bibinfo {author} {\bibfnamefont {R.}~\bibnamefont
  {Hamans}}, \bibinfo {author} {\bibfnamefont {J.-W.}\ \bibnamefont {Liao}},
  \bibinfo {author} {\bibfnamefont {A.}~\bibnamefont {Welbourne}}, \bibinfo
  {author} {\bibfnamefont {R.}~\bibnamefont {Lavrijsen}}, \ and\ \bibinfo
  {author} {\bibfnamefont {A.}~\bibnamefont {{Fern{\'a}ndez-Pacheco}}},\
  }\bibfield  {title} {\enquote {\bibinfo {title} {Fabrication, {{Detection}},
  and {{Operation}} of a {{Three-Dimensional Nanomagnetic Conduit}}},}\
  }\href@noop {} {\bibfield  {journal} {\bibinfo  {journal} {ACS Nano}\
  }\textbf {\bibinfo {volume} {11}} (\bibinfo {year} {2017})}\BibitemShut
  {NoStop}%
\bibitem [{\citenamefont {Luo}\ \emph {et~al.}(2020)\citenamefont {Luo},
  \citenamefont {Hrabec}, \citenamefont {Dao}, \citenamefont {Sala},
  \citenamefont {Finizio}, \citenamefont {Feng}, \citenamefont {Mayr},
  \citenamefont {Raabe}, \citenamefont {Gambardella},\ and\ \citenamefont
  {Heyderman}}]{luo2020current}%
  \BibitemOpen
  \bibfield  {author} {\bibinfo {author} {\bibfnamefont {Z.}~\bibnamefont
  {Luo}}, \bibinfo {author} {\bibfnamefont {A.}~\bibnamefont {Hrabec}},
  \bibinfo {author} {\bibfnamefont {T.~P.}\ \bibnamefont {Dao}}, \bibinfo
  {author} {\bibfnamefont {G.}~\bibnamefont {Sala}}, \bibinfo {author}
  {\bibfnamefont {S.}~\bibnamefont {Finizio}}, \bibinfo {author} {\bibfnamefont
  {J.}~\bibnamefont {Feng}}, \bibinfo {author} {\bibfnamefont {S.}~\bibnamefont
  {Mayr}}, \bibinfo {author} {\bibfnamefont {J.}~\bibnamefont {Raabe}},
  \bibinfo {author} {\bibfnamefont {P.}~\bibnamefont {Gambardella}}, \ and\
  \bibinfo {author} {\bibfnamefont {L.~J.}\ \bibnamefont {Heyderman}},\
  }\bibfield  {title} {\enquote {\bibinfo {title} {Current-driven magnetic
  domain-wall logic},}\ }\href@noop {} {\bibfield  {journal} {\bibinfo
  {journal} {Nature}\ }\textbf {\bibinfo {volume} {579}},\ \bibinfo {pages}
  {214--218} (\bibinfo {year} {2020})}\BibitemShut {NoStop}%
\bibitem [{\citenamefont {Hayward}(2015)}]{hayward2015intrinsic}%
  \BibitemOpen
  \bibfield  {author} {\bibinfo {author} {\bibfnamefont {T.}~\bibnamefont
  {Hayward}},\ }\bibfield  {title} {\enquote {\bibinfo {title} {Intrinsic
  nature of stochastic domain wall pinning phenomena in magnetic nanowire
  devices},}\ }\href@noop {} {\bibfield  {journal} {\bibinfo  {journal}
  {Scientific reports}\ }\textbf {\bibinfo {volume} {5}},\ \bibinfo {pages}
  {1--12} (\bibinfo {year} {2015})}\BibitemShut {NoStop}%
\bibitem [{\citenamefont {Kumar}\ \emph {et~al.}(2019)\citenamefont {Kumar},
  \citenamefont {Jin}, \citenamefont {Al~Risi}, \citenamefont {Sbiaa},
  \citenamefont {Lew},\ and\ \citenamefont
  {Piramanayagam}}]{kumarDomainWallMotion2019}%
  \BibitemOpen
  \bibfield  {author} {\bibinfo {author} {\bibfnamefont {D.}~\bibnamefont
  {Kumar}}, \bibinfo {author} {\bibfnamefont {T.}~\bibnamefont {Jin}}, \bibinfo
  {author} {\bibfnamefont {S.}~\bibnamefont {Al~Risi}}, \bibinfo {author}
  {\bibfnamefont {R.}~\bibnamefont {Sbiaa}}, \bibinfo {author} {\bibfnamefont
  {W.~S.}\ \bibnamefont {Lew}}, \ and\ \bibinfo {author} {\bibfnamefont
  {S.~N.}\ \bibnamefont {Piramanayagam}},\ }\bibfield  {title} {\enquote
  {\bibinfo {title} {Domain {{Wall Motion Control}} for {{Racetrack Memory
  Applications}}},}\ }\href@noop {} {\bibfield  {journal} {\bibinfo  {journal}
  {IEEE Transactions on Magnetics}\ }\textbf {\bibinfo {volume} {55}},\
  \bibinfo {pages} {1--8} (\bibinfo {year} {2019})}\BibitemShut {NoStop}%
\bibitem [{\citenamefont {Lambson}, \citenamefont {Carlton},\ and\
  \citenamefont {Bokor}(2011)}]{lambsonExploringThermodynamicLimits2011}%
  \BibitemOpen
  \bibfield  {author} {\bibinfo {author} {\bibfnamefont {B.}~\bibnamefont
  {Lambson}}, \bibinfo {author} {\bibfnamefont {D.}~\bibnamefont {Carlton}}, \
  and\ \bibinfo {author} {\bibfnamefont {J.}~\bibnamefont {Bokor}},\ }\bibfield
   {title} {\enquote {\bibinfo {title} {Exploring the {{Thermodynamic Limits}}
  of {{Computation}} in {{Integrated Systems}}: {{Magnetic Memory}},
  {{Nanomagnetic Logic}}, and the {{Landauer Limit}}},}\ }\href
  {https://link.aps.org/doi/10.1103/PhysRevLett.107.010604} {\bibfield
  {journal} {\bibinfo  {journal} {Physical Review Letters}\ }\textbf {\bibinfo
  {volume} {107}},\ \bibinfo {pages} {010604} (\bibinfo {year}
  {2011})}\BibitemShut {NoStop}%
\bibitem [{\citenamefont {Hayashi}\ \emph {et~al.}(2008)\citenamefont
  {Hayashi}, \citenamefont {Thomas}, \citenamefont {Moriya}, \citenamefont
  {Rettner},\ and\ \citenamefont
  {Parkin}}]{hayashiCurrentControlledMagneticDomainWall2008}%
  \BibitemOpen
  \bibfield  {author} {\bibinfo {author} {\bibfnamefont {M.}~\bibnamefont
  {Hayashi}}, \bibinfo {author} {\bibfnamefont {L.}~\bibnamefont {Thomas}},
  \bibinfo {author} {\bibfnamefont {R.}~\bibnamefont {Moriya}}, \bibinfo
  {author} {\bibfnamefont {C.}~\bibnamefont {Rettner}}, \ and\ \bibinfo
  {author} {\bibfnamefont {S.~S.~P.}\ \bibnamefont {Parkin}},\ }\bibfield
  {title} {\enquote {\bibinfo {title} {Current-{{Controlled Magnetic
  Domain-Wall Nanowire Shift Register}}},}\ }\href
  {https://www.science.org/doi/10.1126/science.1154587} {\bibfield  {journal}
  {\bibinfo  {journal} {Science}\ }\textbf {\bibinfo {volume} {320}},\ \bibinfo
  {pages} {209--211} (\bibinfo {year} {2008})}\BibitemShut {NoStop}%
\bibitem [{\citenamefont {Yang}, \citenamefont {Ryu},\ and\ \citenamefont
  {Parkin}(2015)}]{Yang2015a}%
  \BibitemOpen
  \bibfield  {author} {\bibinfo {author} {\bibfnamefont {S.-H.}\ \bibnamefont
  {Yang}}, \bibinfo {author} {\bibfnamefont {K.-S.}\ \bibnamefont {Ryu}}, \
  and\ \bibinfo {author} {\bibfnamefont {S.}~\bibnamefont {Parkin}},\
  }\bibfield  {title} {\enquote {\bibinfo {title} {Domain-wall velocities of up
  to 750 m s-1 driven by exchange-coupling torque in synthetic
  antiferromagnets},}\ }\href
  {http://www.nature.com/doifinder/10.1038/nnano.2014.324} {\bibfield
  {journal} {\bibinfo  {journal} {Nature Nanotechnology}\ }\textbf {\bibinfo
  {volume} {10}},\ \bibinfo {pages} {221--226} (\bibinfo {year}
  {2015})}\BibitemShut {NoStop}%
\bibitem [{\citenamefont {Nakatani}, \citenamefont {Thiaville},\ and\
  \citenamefont {Miltat}(2005)}]{nakataniHeadtoheadDomainWalls2005a}%
  \BibitemOpen
  \bibfield  {author} {\bibinfo {author} {\bibfnamefont {Y.}~\bibnamefont
  {Nakatani}}, \bibinfo {author} {\bibfnamefont {A.}~\bibnamefont {Thiaville}},
  \ and\ \bibinfo {author} {\bibfnamefont {J.}~\bibnamefont {Miltat}},\
  }\bibfield  {title} {\enquote {\bibinfo {title} {Head-to-head domain walls in
  soft nano-strips: a refined phase diagram},}\ }\href
  {https://www.sciencedirect.com/science/article/pii/S030488530401618X}
  {\bibfield  {journal} {\bibinfo  {journal} {Journal of Magnetism and Magnetic
  Materials}\ }\bibinfo {series} {Proceedings of the {Joint} {European}
  {Magnetic} {Symposia} ({JEMS}' 04)},\ \textbf {\bibinfo {volume} {290-291}},\
  \bibinfo {pages} {750--753} (\bibinfo {year} {2005})}\BibitemShut {NoStop}%
\bibitem [{\citenamefont {Satel}, \citenamefont {Trappenberg},\ and\
  \citenamefont {Fine}(2009)}]{satel2009binary}%
  \BibitemOpen
  \bibfield  {author} {\bibinfo {author} {\bibfnamefont {J.}~\bibnamefont
  {Satel}}, \bibinfo {author} {\bibfnamefont {T.}~\bibnamefont {Trappenberg}},
  \ and\ \bibinfo {author} {\bibfnamefont {A.}~\bibnamefont {Fine}},\
  }\bibfield  {title} {\enquote {\bibinfo {title} {Are binary synapses superior
  to graded weight representations in stochastic attractor networks?}}\
  }\href@noop {} {\bibfield  {journal} {\bibinfo  {journal} {Cognitive
  neurodynamics}\ }\textbf {\bibinfo {volume} {3}},\ \bibinfo {pages}
  {243--250} (\bibinfo {year} {2009})}\BibitemShut {NoStop}%
\bibitem [{\citenamefont {Dubreuil}, \citenamefont {Amit},\ and\ \citenamefont
  {Brunel}(2014)}]{dubreuil2014memory}%
  \BibitemOpen
  \bibfield  {author} {\bibinfo {author} {\bibfnamefont {A.~M.}\ \bibnamefont
  {Dubreuil}}, \bibinfo {author} {\bibfnamefont {Y.}~\bibnamefont {Amit}}, \
  and\ \bibinfo {author} {\bibfnamefont {N.}~\bibnamefont {Brunel}},\
  }\bibfield  {title} {\enquote {\bibinfo {title} {Memory capacity of networks
  with stochastic binary synapses},}\ }\href@noop {} {\bibfield  {journal}
  {\bibinfo  {journal} {PLoS computational biology}\ }\textbf {\bibinfo
  {volume} {10}},\ \bibinfo {pages} {e1003727} (\bibinfo {year}
  {2014})}\BibitemShut {NoStop}%
\bibitem [{\citenamefont {Hong}(2013)}]{hong2013computing}%
  \BibitemOpen
  \bibfield  {author} {\bibinfo {author} {\bibfnamefont {Y.}~\bibnamefont
  {Hong}},\ }\bibfield  {title} {\enquote {\bibinfo {title} {On computing the
  distribution function for the poisson binomial distribution},}\ }\href@noop
  {} {\bibfield  {journal} {\bibinfo  {journal} {Computational Statistics \&
  Data Analysis}\ }\textbf {\bibinfo {volume} {59}},\ \bibinfo {pages} {41--51}
  (\bibinfo {year} {2013})}\BibitemShut {NoStop}%
\bibitem [{\citenamefont {Vouros}\ \emph {et~al.}(2018)\citenamefont {Vouros},
  \citenamefont {Gehring}, \citenamefont {Szydlowska}, \citenamefont {Janusz},
  \citenamefont {Tu}, \citenamefont {Croucher}, \citenamefont {Lukasiuk},
  \citenamefont {Konopka}, \citenamefont {Sandi},\ and\ \citenamefont
  {Vasilaki}}]{vouros2018generalised}%
  \BibitemOpen
  \bibfield  {author} {\bibinfo {author} {\bibfnamefont {A.}~\bibnamefont
  {Vouros}}, \bibinfo {author} {\bibfnamefont {T.~V.}\ \bibnamefont {Gehring}},
  \bibinfo {author} {\bibfnamefont {K.}~\bibnamefont {Szydlowska}}, \bibinfo
  {author} {\bibfnamefont {A.}~\bibnamefont {Janusz}}, \bibinfo {author}
  {\bibfnamefont {Z.}~\bibnamefont {Tu}}, \bibinfo {author} {\bibfnamefont
  {M.}~\bibnamefont {Croucher}}, \bibinfo {author} {\bibfnamefont
  {K.}~\bibnamefont {Lukasiuk}}, \bibinfo {author} {\bibfnamefont
  {W.}~\bibnamefont {Konopka}}, \bibinfo {author} {\bibfnamefont
  {C.}~\bibnamefont {Sandi}}, \ and\ \bibinfo {author} {\bibfnamefont
  {E.}~\bibnamefont {Vasilaki}},\ }\bibfield  {title} {\enquote {\bibinfo
  {title} {A generalised framework for detailed classification of swimming
  paths inside the morris water maze},}\ }\href@noop {} {\bibfield  {journal}
  {\bibinfo  {journal} {Scientific reports}\ }\textbf {\bibinfo {volume} {8}},\
  \bibinfo {pages} {1--15} (\bibinfo {year} {2018})}\BibitemShut {NoStop}%
\bibitem [{\citenamefont {Esser}\ \emph {et~al.}(2015)\citenamefont {Esser},
  \citenamefont {Appuswamy}, \citenamefont {Merolla}, \citenamefont {Arthur},\
  and\ \citenamefont {Modha}}]{esser2015backpropagation}%
  \BibitemOpen
  \bibfield  {author} {\bibinfo {author} {\bibfnamefont {S.~K.}\ \bibnamefont
  {Esser}}, \bibinfo {author} {\bibfnamefont {R.}~\bibnamefont {Appuswamy}},
  \bibinfo {author} {\bibfnamefont {P.}~\bibnamefont {Merolla}}, \bibinfo
  {author} {\bibfnamefont {J.~V.}\ \bibnamefont {Arthur}}, \ and\ \bibinfo
  {author} {\bibfnamefont {D.~S.}\ \bibnamefont {Modha}},\ }\bibfield  {title}
  {\enquote {\bibinfo {title} {Backpropagation for energy-efficient
  neuromorphic computing},}\ }\href@noop {} {\bibfield  {journal} {\bibinfo
  {journal} {Advances in neural information processing systems}\ }\textbf
  {\bibinfo {volume} {28}} (\bibinfo {year} {2015})}\BibitemShut {NoStop}%
\bibitem [{\citenamefont {Williams}(1992)}]{williams1992simple}%
  \BibitemOpen
  \bibfield  {author} {\bibinfo {author} {\bibfnamefont {R.~J.}\ \bibnamefont
  {Williams}},\ }\bibfield  {title} {\enquote {\bibinfo {title} {Simple
  statistical gradient-following algorithms for connectionist reinforcement
  learning},}\ }\href@noop {} {\bibfield  {journal} {\bibinfo  {journal}
  {Machine learning}\ }\textbf {\bibinfo {volume} {8}},\ \bibinfo {pages}
  {229--256} (\bibinfo {year} {1992})}\BibitemShut {NoStop}%
\bibitem [{\citenamefont {Gu}\ \emph {et~al.}(2015)\citenamefont {Gu},
  \citenamefont {Levine}, \citenamefont {Sutskever},\ and\ \citenamefont
  {Mnih}}]{gu2015muprop}%
  \BibitemOpen
  \bibfield  {author} {\bibinfo {author} {\bibfnamefont {S.}~\bibnamefont
  {Gu}}, \bibinfo {author} {\bibfnamefont {S.}~\bibnamefont {Levine}}, \bibinfo
  {author} {\bibfnamefont {I.}~\bibnamefont {Sutskever}}, \ and\ \bibinfo
  {author} {\bibfnamefont {A.}~\bibnamefont {Mnih}},\ }\bibfield  {title}
  {\enquote {\bibinfo {title} {Muprop: Unbiased backpropagation for stochastic
  neural networks},}\ }\href@noop {} {\bibfield  {journal} {\bibinfo  {journal}
  {arXiv preprint arXiv:1511.05176}\ } (\bibinfo {year} {2015})}\BibitemShut
  {NoStop}%
\bibitem [{\citenamefont {Parmas}\ and\ \citenamefont
  {Sugiyama}(2021)}]{parmas2021unified}%
  \BibitemOpen
  \bibfield  {author} {\bibinfo {author} {\bibfnamefont {P.}~\bibnamefont
  {Parmas}}\ and\ \bibinfo {author} {\bibfnamefont {M.}~\bibnamefont
  {Sugiyama}},\ }\bibfield  {title} {\enquote {\bibinfo {title} {A unified view
  of likelihood ratio and reparameterization gradients},}\ }in\ \href@noop {}
  {\emph {\bibinfo {booktitle} {International Conference on Artificial
  Intelligence and Statistics}}}\ (\bibinfo {organization} {PMLR},\ \bibinfo
  {year} {2021})\ pp.\ \bibinfo {pages} {4078--4086}\BibitemShut {NoStop}%
\bibitem [{\citenamefont {Vasilaki}\ \emph {et~al.}(2009)\citenamefont
  {Vasilaki}, \citenamefont {Fr{\'e}maux}, \citenamefont {Urbanczik},
  \citenamefont {Senn},\ and\ \citenamefont {Gerstner}}]{vasilaki2009spike}%
  \BibitemOpen
  \bibfield  {author} {\bibinfo {author} {\bibfnamefont {E.}~\bibnamefont
  {Vasilaki}}, \bibinfo {author} {\bibfnamefont {N.}~\bibnamefont
  {Fr{\'e}maux}}, \bibinfo {author} {\bibfnamefont {R.}~\bibnamefont
  {Urbanczik}}, \bibinfo {author} {\bibfnamefont {W.}~\bibnamefont {Senn}}, \
  and\ \bibinfo {author} {\bibfnamefont {W.}~\bibnamefont {Gerstner}},\
  }\bibfield  {title} {\enquote {\bibinfo {title} {Spike-based reinforcement
  learning in continuous state and action space: when policy gradient methods
  fail},}\ }\href@noop {} {\bibfield  {journal} {\bibinfo  {journal} {PLoS
  computational biology}\ }\textbf {\bibinfo {volume} {5}},\ \bibinfo {pages}
  {e1000586} (\bibinfo {year} {2009})}\BibitemShut {NoStop}%
\bibitem [{\citenamefont {Azam}\ \emph {et~al.}(2020)\citenamefont {Azam},
  \citenamefont {Bhattacharya}, \citenamefont {Querlioz}, \citenamefont
  {Ross},\ and\ \citenamefont {Atulasimha}}]{azam2020voltage}%
  \BibitemOpen
  \bibfield  {author} {\bibinfo {author} {\bibfnamefont {M.~A.}\ \bibnamefont
  {Azam}}, \bibinfo {author} {\bibfnamefont {D.}~\bibnamefont {Bhattacharya}},
  \bibinfo {author} {\bibfnamefont {D.}~\bibnamefont {Querlioz}}, \bibinfo
  {author} {\bibfnamefont {C.~A.}\ \bibnamefont {Ross}}, \ and\ \bibinfo
  {author} {\bibfnamefont {J.}~\bibnamefont {Atulasimha}},\ }\bibfield  {title}
  {\enquote {\bibinfo {title} {Voltage control of domain walls in magnetic
  nanowires for energy-efficient neuromorphic devices},}\ }\href@noop {}
  {\bibfield  {journal} {\bibinfo  {journal} {Nanotechnology}\ }\textbf
  {\bibinfo {volume} {31}},\ \bibinfo {pages} {145201} (\bibinfo {year}
  {2020})}\BibitemShut {NoStop}%
\bibitem [{\citenamefont {{Sanz-Hern{\'a}ndez}}\ \emph
  {et~al.}(2021)\citenamefont {{Sanz-Hern{\'a}ndez}}, \citenamefont
  {Massouras}, \citenamefont {Reyren}, \citenamefont {Rougemaille},
  \citenamefont {Sch{\'a}nilec}, \citenamefont {Bouzehouane}, \citenamefont
  {Hehn}, \citenamefont {Canals}, \citenamefont {Querlioz}, \citenamefont
  {Grollier}, \citenamefont {Montaigne},\ and\ \citenamefont
  {Lacour}}]{sanz-hernandezTunableStochasticityArtificial2021}%
  \BibitemOpen
  \bibfield  {author} {\bibinfo {author} {\bibfnamefont {D.}~\bibnamefont
  {{Sanz-Hern{\'a}ndez}}}, \bibinfo {author} {\bibfnamefont {M.}~\bibnamefont
  {Massouras}}, \bibinfo {author} {\bibfnamefont {N.}~\bibnamefont {Reyren}},
  \bibinfo {author} {\bibfnamefont {N.}~\bibnamefont {Rougemaille}}, \bibinfo
  {author} {\bibfnamefont {V.}~\bibnamefont {Sch{\'a}nilec}}, \bibinfo {author}
  {\bibfnamefont {K.}~\bibnamefont {Bouzehouane}}, \bibinfo {author}
  {\bibfnamefont {M.}~\bibnamefont {Hehn}}, \bibinfo {author} {\bibfnamefont
  {B.}~\bibnamefont {Canals}}, \bibinfo {author} {\bibfnamefont
  {D.}~\bibnamefont {Querlioz}}, \bibinfo {author} {\bibfnamefont
  {J.}~\bibnamefont {Grollier}}, \bibinfo {author} {\bibfnamefont
  {F.}~\bibnamefont {Montaigne}}, \ and\ \bibinfo {author} {\bibfnamefont
  {D.}~\bibnamefont {Lacour}},\ }\bibfield  {title} {\enquote {\bibinfo {title}
  {Tunable {{Stochasticity}} in an {{Artificial Spin Network}}},}\ }\href@noop
  {} {\bibfield  {journal} {\bibinfo  {journal} {Advanced Materials}\ }\textbf
  {\bibinfo {volume} {33}},\ \bibinfo {pages} {2008135} (\bibinfo {year}
  {2021})}\BibitemShut {NoStop}%
\bibitem [{\citenamefont {Misba}\ \emph {et~al.}(2022)\citenamefont {Misba},
  \citenamefont {Lozano}, \citenamefont {Querlioz},\ and\ \citenamefont
  {Atulasimha}}]{misbaEnergyEfficientLearning2022}%
  \BibitemOpen
  \bibfield  {author} {\bibinfo {author} {\bibfnamefont {W.~A.}\ \bibnamefont
  {Misba}}, \bibinfo {author} {\bibfnamefont {M.}~\bibnamefont {Lozano}},
  \bibinfo {author} {\bibfnamefont {D.}~\bibnamefont {Querlioz}}, \ and\
  \bibinfo {author} {\bibfnamefont {J.}~\bibnamefont {Atulasimha}},\ }\bibfield
   {title} {\enquote {\bibinfo {title} {Energy {{Efficient Learning With Low
  Resolution Stochastic Domain Wall Synapse}} for {{Deep Neural Networks}}},}\
  }\href@noop {} {\bibfield  {journal} {\bibinfo  {journal} {IEEE Access}\
  }\textbf {\bibinfo {volume} {10}},\ \bibinfo {pages} {84946--84959} (\bibinfo
  {year} {2022})}\BibitemShut {NoStop}%
\bibitem [{\citenamefont {Hassan}\ \emph {et~al.}(2018)\citenamefont {Hassan},
  \citenamefont {Hu}, \citenamefont {Jiang-Wei}, \citenamefont {Brigner},
  \citenamefont {Akinola}, \citenamefont {Garcia-Sanchez}, \citenamefont
  {Pasquale}, \citenamefont {Bennett}, \citenamefont {Incorvia},\ and\
  \citenamefont {Friedman}}]{hassan2018magnetic}%
  \BibitemOpen
  \bibfield  {author} {\bibinfo {author} {\bibfnamefont {N.}~\bibnamefont
  {Hassan}}, \bibinfo {author} {\bibfnamefont {X.}~\bibnamefont {Hu}}, \bibinfo
  {author} {\bibfnamefont {L.}~\bibnamefont {Jiang-Wei}}, \bibinfo {author}
  {\bibfnamefont {W.~H.}\ \bibnamefont {Brigner}}, \bibinfo {author}
  {\bibfnamefont {O.~G.}\ \bibnamefont {Akinola}}, \bibinfo {author}
  {\bibfnamefont {F.}~\bibnamefont {Garcia-Sanchez}}, \bibinfo {author}
  {\bibfnamefont {M.}~\bibnamefont {Pasquale}}, \bibinfo {author}
  {\bibfnamefont {C.~H.}\ \bibnamefont {Bennett}}, \bibinfo {author}
  {\bibfnamefont {J.~A.~C.}\ \bibnamefont {Incorvia}}, \ and\ \bibinfo {author}
  {\bibfnamefont {J.~S.}\ \bibnamefont {Friedman}},\ }\bibfield  {title}
  {\enquote {\bibinfo {title} {Magnetic domain wall neuron with lateral
  inhibition},}\ }\href@noop {} {\bibfield  {journal} {\bibinfo  {journal}
  {Journal of Applied Physics}\ }\textbf {\bibinfo {volume} {124}},\ \bibinfo
  {pages} {152127} (\bibinfo {year} {2018})}\BibitemShut {NoStop}%
\bibitem [{\citenamefont {Brigner}\ \emph {et~al.}(2022)\citenamefont
  {Brigner}, \citenamefont {Hassan}, \citenamefont {Hu}, \citenamefont
  {Bennett}, \citenamefont {Garcia-Sanchez}, \citenamefont {Cui}, \citenamefont
  {Velasquez}, \citenamefont {Marinella}, \citenamefont {Incorvia},\ and\
  \citenamefont {Friedman}}]{brigner2022domain}%
  \BibitemOpen
  \bibfield  {author} {\bibinfo {author} {\bibfnamefont {W.~H.}\ \bibnamefont
  {Brigner}}, \bibinfo {author} {\bibfnamefont {N.}~\bibnamefont {Hassan}},
  \bibinfo {author} {\bibfnamefont {X.}~\bibnamefont {Hu}}, \bibinfo {author}
  {\bibfnamefont {C.~H.}\ \bibnamefont {Bennett}}, \bibinfo {author}
  {\bibfnamefont {F.}~\bibnamefont {Garcia-Sanchez}}, \bibinfo {author}
  {\bibfnamefont {C.}~\bibnamefont {Cui}}, \bibinfo {author} {\bibfnamefont
  {A.}~\bibnamefont {Velasquez}}, \bibinfo {author} {\bibfnamefont {M.~J.}\
  \bibnamefont {Marinella}}, \bibinfo {author} {\bibfnamefont {J.~A.~C.}\
  \bibnamefont {Incorvia}}, \ and\ \bibinfo {author} {\bibfnamefont {J.~S.}\
  \bibnamefont {Friedman}},\ }\bibfield  {title} {\enquote {\bibinfo {title}
  {Domain wall leaky integrate-and-fire neurons with shape-based configurable
  activation functions},}\ }\href@noop {} {\bibfield  {journal} {\bibinfo
  {journal} {IEEE Transactions on Electron Devices}\ }\textbf {\bibinfo
  {volume} {69}},\ \bibinfo {pages} {2353--2359} (\bibinfo {year}
  {2022})}\BibitemShut {NoStop}%
\bibitem [{\citenamefont {Borders}\ \emph {et~al.}(2019)\citenamefont
  {Borders}, \citenamefont {Pervaiz}, \citenamefont {Fukami}, \citenamefont
  {Camsari}, \citenamefont {Ohno},\ and\ \citenamefont
  {Datta}}]{borders2019integer}%
  \BibitemOpen
  \bibfield  {author} {\bibinfo {author} {\bibfnamefont {W.~A.}\ \bibnamefont
  {Borders}}, \bibinfo {author} {\bibfnamefont {A.~Z.}\ \bibnamefont
  {Pervaiz}}, \bibinfo {author} {\bibfnamefont {S.}~\bibnamefont {Fukami}},
  \bibinfo {author} {\bibfnamefont {K.~Y.}\ \bibnamefont {Camsari}}, \bibinfo
  {author} {\bibfnamefont {H.}~\bibnamefont {Ohno}}, \ and\ \bibinfo {author}
  {\bibfnamefont {S.}~\bibnamefont {Datta}},\ }\bibfield  {title} {\enquote
  {\bibinfo {title} {Integer factorization using stochastic magnetic tunnel
  junctions},}\ }\href@noop {} {\bibfield  {journal} {\bibinfo  {journal}
  {Nature}\ }\textbf {\bibinfo {volume} {573}},\ \bibinfo {pages} {390--393}
  (\bibinfo {year} {2019})}\BibitemShut {NoStop}%
\bibitem [{\citenamefont {Al~Misba}\ \emph {et~al.}(2022)\citenamefont
  {Al~Misba}, \citenamefont {Lozano}, \citenamefont {Querlioz},\ and\
  \citenamefont {Atulasimha}}]{al2022energy}%
  \BibitemOpen
  \bibfield  {author} {\bibinfo {author} {\bibfnamefont {W.}~\bibnamefont
  {Al~Misba}}, \bibinfo {author} {\bibfnamefont {M.}~\bibnamefont {Lozano}},
  \bibinfo {author} {\bibfnamefont {D.}~\bibnamefont {Querlioz}}, \ and\
  \bibinfo {author} {\bibfnamefont {J.}~\bibnamefont {Atulasimha}},\ }\bibfield
   {title} {\enquote {\bibinfo {title} {Energy efficient learning with low
  resolution stochastic domain wall synapse for deep neural networks},}\
  }\href@noop {} {\bibfield  {journal} {\bibinfo  {journal} {IEEE Access}\
  }\textbf {\bibinfo {volume} {10}},\ \bibinfo {pages} {84946--84959} (\bibinfo
  {year} {2022})}\BibitemShut {NoStop}%
\bibitem [{\citenamefont {Koo}\ \emph {et~al.}(2020)\citenamefont {Koo},
  \citenamefont {Srinivasan}, \citenamefont {Shim},\ and\ \citenamefont
  {Roy}}]{koo2020sbsnn}%
  \BibitemOpen
  \bibfield  {author} {\bibinfo {author} {\bibfnamefont {M.}~\bibnamefont
  {Koo}}, \bibinfo {author} {\bibfnamefont {G.}~\bibnamefont {Srinivasan}},
  \bibinfo {author} {\bibfnamefont {Y.}~\bibnamefont {Shim}}, \ and\ \bibinfo
  {author} {\bibfnamefont {K.}~\bibnamefont {Roy}},\ }\bibfield  {title}
  {\enquote {\bibinfo {title} {Sbsnn: Stochastic-bits enabled binary spiking
  neural network with on-chip learning for energy efficient neuromorphic
  computing at the edge},}\ }\href@noop {} {\bibfield  {journal} {\bibinfo
  {journal} {IEEE Transactions on Circuits and Systems I: Regular Papers}\
  }\textbf {\bibinfo {volume} {67}},\ \bibinfo {pages} {2546--2555} (\bibinfo
  {year} {2020})}\BibitemShut {NoStop}%
\bibitem [{\citenamefont {Dawidek}\ \emph {et~al.}(2021)\citenamefont
  {Dawidek}, \citenamefont {Hayward}, \citenamefont {Vidamour}, \citenamefont
  {Broomhall}, \citenamefont {Venkat}, \citenamefont {Mamoori}, \citenamefont
  {Mullen}, \citenamefont {Kyle}, \citenamefont {Fry}, \citenamefont {Steinke},
  \citenamefont {Cooper}, \citenamefont {Maccherozzi}, \citenamefont {Dhesi},
  \citenamefont {Aballe}, \citenamefont {Foerster}, \citenamefont {Prat},
  \citenamefont {Vasilaki}, \citenamefont {Ellis},\ and\ \citenamefont
  {Allwood}}]{Dawidek2021}%
  \BibitemOpen
  \bibfield  {author} {\bibinfo {author} {\bibfnamefont {R.~W.}\ \bibnamefont
  {Dawidek}}, \bibinfo {author} {\bibfnamefont {T.~J.}\ \bibnamefont
  {Hayward}}, \bibinfo {author} {\bibfnamefont {I.~T.}\ \bibnamefont
  {Vidamour}}, \bibinfo {author} {\bibfnamefont {T.~J.}\ \bibnamefont
  {Broomhall}}, \bibinfo {author} {\bibfnamefont {G.}~\bibnamefont {Venkat}},
  \bibinfo {author} {\bibfnamefont {M.~A.}\ \bibnamefont {Mamoori}}, \bibinfo
  {author} {\bibfnamefont {A.}~\bibnamefont {Mullen}}, \bibinfo {author}
  {\bibfnamefont {S.~J.}\ \bibnamefont {Kyle}}, \bibinfo {author}
  {\bibfnamefont {P.~W.}\ \bibnamefont {Fry}}, \bibinfo {author} {\bibfnamefont
  {N.-J.}\ \bibnamefont {Steinke}}, \bibinfo {author} {\bibfnamefont
  {J.~F.~K.}\ \bibnamefont {Cooper}}, \bibinfo {author} {\bibfnamefont
  {F.}~\bibnamefont {Maccherozzi}}, \bibinfo {author} {\bibfnamefont {S.~S.}\
  \bibnamefont {Dhesi}}, \bibinfo {author} {\bibfnamefont {L.}~\bibnamefont
  {Aballe}}, \bibinfo {author} {\bibfnamefont {M.}~\bibnamefont {Foerster}},
  \bibinfo {author} {\bibfnamefont {J.}~\bibnamefont {Prat}}, \bibinfo {author}
  {\bibfnamefont {E.}~\bibnamefont {Vasilaki}}, \bibinfo {author}
  {\bibfnamefont {M.~O.~A.}\ \bibnamefont {Ellis}}, \ and\ \bibinfo {author}
  {\bibfnamefont {D.~A.}\ \bibnamefont {Allwood}},\ }\bibfield  {title}
  {\enquote {\bibinfo {title} {Dynamically-driven emergence in a nanomagnetic
  system},}\ }\href
  {https://onlinelibrary.wiley.com/doi/abs/10.1002/adfm.202008389} {\bibfield
  {journal} {\bibinfo  {journal} {Advanced Functional Materials}\ }\textbf
  {\bibinfo {volume} {31}},\ \bibinfo {pages} {2008389} (\bibinfo {year}
  {2021})}\BibitemShut {NoStop}%
\bibitem [{\citenamefont {Ababei}\ \emph {et~al.}(2021)\citenamefont {Ababei},
  \citenamefont {Ellis}, \citenamefont {Vidamour}, \citenamefont {Devadasan},
  \citenamefont {Allwood}, \citenamefont {Vasilaki},\ and\ \citenamefont
  {Hayward}}]{Ababei2021}%
  \BibitemOpen
  \bibfield  {author} {\bibinfo {author} {\bibfnamefont {R.~V.}\ \bibnamefont
  {Ababei}}, \bibinfo {author} {\bibfnamefont {M.~O.~A.}\ \bibnamefont
  {Ellis}}, \bibinfo {author} {\bibfnamefont {I.~T.}\ \bibnamefont {Vidamour}},
  \bibinfo {author} {\bibfnamefont {D.~S.}\ \bibnamefont {Devadasan}}, \bibinfo
  {author} {\bibfnamefont {D.~A.}\ \bibnamefont {Allwood}}, \bibinfo {author}
  {\bibfnamefont {E.}~\bibnamefont {Vasilaki}}, \ and\ \bibinfo {author}
  {\bibfnamefont {T.~J.}\ \bibnamefont {Hayward}},\ }\bibfield  {title}
  {\enquote {\bibinfo {title} {Neuromorphic computation with a single magnetic
  domain wall},}\ }\href {https://doi.org/10.1038/s41598-021-94975-y}
  {\bibfield  {journal} {\bibinfo  {journal} {Scientific Reports}\ }\textbf
  {\bibinfo {volume} {11}},\ \bibinfo {pages} {15587} (\bibinfo {year}
  {2021})}\BibitemShut {NoStop}%
\bibitem [{\citenamefont {Welbourne}\ \emph {et~al.}(2021)\citenamefont
  {Welbourne}, \citenamefont {Levy}, \citenamefont {Ellis}, \citenamefont
  {Chen}, \citenamefont {Thompson}, \citenamefont {Vasilaki}, \citenamefont
  {Allwood},\ and\ \citenamefont {Hayward}}]{Welbourne2021}%
  \BibitemOpen
  \bibfield  {author} {\bibinfo {author} {\bibfnamefont {A.}~\bibnamefont
  {Welbourne}}, \bibinfo {author} {\bibfnamefont {A.~L.~R.}\ \bibnamefont
  {Levy}}, \bibinfo {author} {\bibfnamefont {M.~O.~A.}\ \bibnamefont {Ellis}},
  \bibinfo {author} {\bibfnamefont {H.}~\bibnamefont {Chen}}, \bibinfo {author}
  {\bibfnamefont {M.~J.}\ \bibnamefont {Thompson}}, \bibinfo {author}
  {\bibfnamefont {E.}~\bibnamefont {Vasilaki}}, \bibinfo {author}
  {\bibfnamefont {D.~A.}\ \bibnamefont {Allwood}}, \ and\ \bibinfo {author}
  {\bibfnamefont {T.~J.}\ \bibnamefont {Hayward}},\ }\bibfield  {title}
  {\enquote {\bibinfo {title} {Voltage-controlled superparamagnetic ensembles
  for low-power reservoir computing},}\ }\href
  {https://doi.org/10.1063/5.0048911} {\bibfield  {journal} {\bibinfo
  {journal} {Applied Physics Letters}\ }\textbf {\bibinfo {volume} {118}},\
  \bibinfo {pages} {202402} (\bibinfo {year} {2021})}\BibitemShut {NoStop}%
\bibitem [{\citenamefont {Gartside}\ \emph {et~al.}(2022)\citenamefont
  {Gartside}, \citenamefont {Stenning}, \citenamefont {Vanstone}, \citenamefont
  {Holder}, \citenamefont {Arroo}, \citenamefont {Dion}, \citenamefont
  {Caravelli}, \citenamefont {Kurebayashi},\ and\ \citenamefont
  {Branford}}]{gartsideReconfigurableTrainingReservoir2022}%
  \BibitemOpen
  \bibfield  {author} {\bibinfo {author} {\bibfnamefont {J.~C.}\ \bibnamefont
  {Gartside}}, \bibinfo {author} {\bibfnamefont {K.~D.}\ \bibnamefont
  {Stenning}}, \bibinfo {author} {\bibfnamefont {A.}~\bibnamefont {Vanstone}},
  \bibinfo {author} {\bibfnamefont {H.~H.}\ \bibnamefont {Holder}}, \bibinfo
  {author} {\bibfnamefont {D.~M.}\ \bibnamefont {Arroo}}, \bibinfo {author}
  {\bibfnamefont {T.}~\bibnamefont {Dion}}, \bibinfo {author} {\bibfnamefont
  {F.}~\bibnamefont {Caravelli}}, \bibinfo {author} {\bibfnamefont
  {H.}~\bibnamefont {Kurebayashi}}, \ and\ \bibinfo {author} {\bibfnamefont
  {W.~R.}\ \bibnamefont {Branford}},\ }\bibfield  {title} {\enquote {\bibinfo
  {title} {Reconfigurable training and reservoir computing in an artificial
  spin-vortex ice via spin-wave fingerprinting},}\ }\href@noop {} {\bibfield
  {journal} {\bibinfo  {journal} {Nature Nanotechnology}\ }\textbf {\bibinfo
  {volume} {17}},\ \bibinfo {pages} {460--469} (\bibinfo {year}
  {2022})}\BibitemShut {NoStop}%
\bibitem [{\citenamefont {Vidamour}\ \emph
  {et~al.}(2022{\natexlab{a}})\citenamefont {Vidamour}, \citenamefont {Ellis},
  \citenamefont {Griffin}, \citenamefont {Venkat}, \citenamefont {Swindells},
  \citenamefont {Dawidek}, \citenamefont {Broomhall}, \citenamefont {Steinke},
  \citenamefont {Cooper}, \citenamefont {Maccherozzi}, \citenamefont {Dhesi},
  \citenamefont {Stepney}, \citenamefont {Vasilaki}, \citenamefont {Allwood},\
  and\ \citenamefont
  {Hayward}}]{vidamourQuantifyingComputationalCapability2022}%
  \BibitemOpen
  \bibfield  {author} {\bibinfo {author} {\bibfnamefont {I.~T.}\ \bibnamefont
  {Vidamour}}, \bibinfo {author} {\bibfnamefont {M.~O.~A.}\ \bibnamefont
  {Ellis}}, \bibinfo {author} {\bibfnamefont {D.}~\bibnamefont {Griffin}},
  \bibinfo {author} {\bibfnamefont {G.}~\bibnamefont {Venkat}}, \bibinfo
  {author} {\bibfnamefont {C.}~\bibnamefont {Swindells}}, \bibinfo {author}
  {\bibfnamefont {R.~W.~S.}\ \bibnamefont {Dawidek}}, \bibinfo {author}
  {\bibfnamefont {T.~J.}\ \bibnamefont {Broomhall}}, \bibinfo {author}
  {\bibfnamefont {N.~J.}\ \bibnamefont {Steinke}}, \bibinfo {author}
  {\bibfnamefont {J.~F.~K.}\ \bibnamefont {Cooper}}, \bibinfo {author}
  {\bibfnamefont {F.}~\bibnamefont {Maccherozzi}}, \bibinfo {author}
  {\bibfnamefont {S.~S.}\ \bibnamefont {Dhesi}}, \bibinfo {author}
  {\bibfnamefont {S.}~\bibnamefont {Stepney}}, \bibinfo {author} {\bibfnamefont
  {E.}~\bibnamefont {Vasilaki}}, \bibinfo {author} {\bibfnamefont {D.~A.}\
  \bibnamefont {Allwood}}, \ and\ \bibinfo {author} {\bibfnamefont {T.~J.}\
  \bibnamefont {Hayward}},\ }\bibfield  {title} {\enquote {\bibinfo {title}
  {Quantifying the computational capability of a nanomagnetic reservoir
  computing platform with emergent magnetisation dynamics},}\ }\href@noop {}
  {\bibfield  {journal} {\bibinfo  {journal} {Nanotechnology}\ }\textbf
  {\bibinfo {volume} {33}},\ \bibinfo {pages} {485203} (\bibinfo {year}
  {2022}{\natexlab{a}})}\BibitemShut {NoStop}%
\bibitem [{\citenamefont {Vidamour}\ \emph
  {et~al.}(2022{\natexlab{b}})\citenamefont {Vidamour}, \citenamefont
  {Swindells}, \citenamefont {Venkat}, \citenamefont {Fry}, \citenamefont
  {Welbourne}, \citenamefont {{Rowan-Robinson}}, \citenamefont {Backes},
  \citenamefont {Maccherozzi}, \citenamefont {Dhesi}, \citenamefont {Vasilaki},
  \citenamefont {Allwood},\ and\ \citenamefont
  {Hayward}}]{vidamourReservoirComputingEmergent2022}%
  \BibitemOpen
  \bibfield  {author} {\bibinfo {author} {\bibfnamefont {I.}~\bibnamefont
  {Vidamour}}, \bibinfo {author} {\bibfnamefont {C.}~\bibnamefont {Swindells}},
  \bibinfo {author} {\bibfnamefont {G.}~\bibnamefont {Venkat}}, \bibinfo
  {author} {\bibfnamefont {P.}~\bibnamefont {Fry}}, \bibinfo {author}
  {\bibfnamefont {A.}~\bibnamefont {Welbourne}}, \bibinfo {author}
  {\bibfnamefont {R.}~\bibnamefont {{Rowan-Robinson}}}, \bibinfo {author}
  {\bibfnamefont {D.}~\bibnamefont {Backes}}, \bibinfo {author} {\bibfnamefont
  {F.}~\bibnamefont {Maccherozzi}}, \bibinfo {author} {\bibfnamefont
  {S.}~\bibnamefont {Dhesi}}, \bibinfo {author} {\bibfnamefont
  {E.}~\bibnamefont {Vasilaki}}, \bibinfo {author} {\bibfnamefont
  {D.}~\bibnamefont {Allwood}}, \ and\ \bibinfo {author} {\bibfnamefont
  {T.}~\bibnamefont {Hayward}},\ }\href {http://arxiv.org/abs/2206.04446}
  {\enquote {\bibinfo {title} {Reservoir {{Computing}} with {{Emergent
  Dynamics}} in a {{Magnetic Metamaterial}}},}\ } (\bibinfo {year}
  {2022}{\natexlab{b}}),\ \Eprint {http://arxiv.org/abs/2206.04446}
  {arXiv:2206.04446 [cond-mat]} \BibitemShut {NoStop}%
\bibitem [{\citenamefont {Allwood}\ \emph {et~al.}(2022)\citenamefont
  {Allwood}, \citenamefont {Ellis}, \citenamefont {Griffin}, \citenamefont
  {Hayward}, \citenamefont {Manneschi}, \citenamefont {Musameh}, \citenamefont
  {O'Keefe}, \citenamefont {Stepney}, \citenamefont {Swindells}, \citenamefont
  {Trefzer}, \citenamefont {Vasilaki}, \citenamefont {Venkat}, \citenamefont
  {Vidamour},\ and\ \citenamefont
  {Wringe}}]{allwoodPerspectivePhysicalReservoir2022}%
  \BibitemOpen
  \bibfield  {author} {\bibinfo {author} {\bibfnamefont {D.~A.}\ \bibnamefont
  {Allwood}}, \bibinfo {author} {\bibfnamefont {M.~O.~A.}\ \bibnamefont
  {Ellis}}, \bibinfo {author} {\bibfnamefont {D.}~\bibnamefont {Griffin}},
  \bibinfo {author} {\bibfnamefont {T.~J.}\ \bibnamefont {Hayward}}, \bibinfo
  {author} {\bibfnamefont {L.}~\bibnamefont {Manneschi}}, \bibinfo {author}
  {\bibfnamefont {M.~F.~K.}\ \bibnamefont {Musameh}}, \bibinfo {author}
  {\bibfnamefont {S.}~\bibnamefont {O'Keefe}}, \bibinfo {author} {\bibfnamefont
  {S.}~\bibnamefont {Stepney}}, \bibinfo {author} {\bibfnamefont
  {C.}~\bibnamefont {Swindells}}, \bibinfo {author} {\bibfnamefont {M.~A.}\
  \bibnamefont {Trefzer}}, \bibinfo {author} {\bibfnamefont {E.}~\bibnamefont
  {Vasilaki}}, \bibinfo {author} {\bibfnamefont {G.}~\bibnamefont {Venkat}},
  \bibinfo {author} {\bibfnamefont {I.}~\bibnamefont {Vidamour}}, \ and\
  \bibinfo {author} {\bibfnamefont {C.}~\bibnamefont {Wringe}},\ }\href
  {http://arxiv.org/abs/2212.04851} {\enquote {\bibinfo {title} {A perspective
  on physical reservoir computing with nanomagnetic devices},}\ } (\bibinfo
  {year} {2022}),\ \Eprint {http://arxiv.org/abs/2212.04851} {arXiv:2212.04851
  [physics]} \BibitemShut {NoStop}%
\bibitem [{\citenamefont {Stenning}\ \emph {et~al.}(2022)\citenamefont
  {Stenning}, \citenamefont {Gartside}, \citenamefont {Manneschi},
  \citenamefont {Cheung}, \citenamefont {Chen}, \citenamefont {Vanstone},
  \citenamefont {Love}, \citenamefont {Holder}, \citenamefont {Caravelli},
  \citenamefont {{Everschor-Sitte}}, \citenamefont {Vasilaki},\ and\
  \citenamefont {Branford}}]{stenningAdaptiveProgrammableNetworks2022}%
  \BibitemOpen
  \bibfield  {author} {\bibinfo {author} {\bibfnamefont {K.~D.}\ \bibnamefont
  {Stenning}}, \bibinfo {author} {\bibfnamefont {J.~C.}\ \bibnamefont
  {Gartside}}, \bibinfo {author} {\bibfnamefont {L.}~\bibnamefont {Manneschi}},
  \bibinfo {author} {\bibfnamefont {C.~T.~S.}\ \bibnamefont {Cheung}}, \bibinfo
  {author} {\bibfnamefont {T.}~\bibnamefont {Chen}}, \bibinfo {author}
  {\bibfnamefont {A.}~\bibnamefont {Vanstone}}, \bibinfo {author}
  {\bibfnamefont {J.}~\bibnamefont {Love}}, \bibinfo {author} {\bibfnamefont
  {H.~H.}\ \bibnamefont {Holder}}, \bibinfo {author} {\bibfnamefont
  {F.}~\bibnamefont {Caravelli}}, \bibinfo {author} {\bibfnamefont
  {K.}~\bibnamefont {{Everschor-Sitte}}}, \bibinfo {author} {\bibfnamefont
  {E.}~\bibnamefont {Vasilaki}}, \ and\ \bibinfo {author} {\bibfnamefont
  {W.~R.}\ \bibnamefont {Branford}},\ }\href {http://arxiv.org/abs/2211.06373}
  {\enquote {\bibinfo {title} {Adaptive {{Programmable Networks}} for {{In
  Materia Neuromorphic Computing}}},}\ } (\bibinfo {year} {2022}),\ \Eprint
  {http://arxiv.org/abs/2211.06373} {arXiv:2211.06373 [cond-mat]} \BibitemShut
  {NoStop}%
\bibitem [{\citenamefont {Hirtzlin}\ \emph {et~al.}(2020)\citenamefont
  {Hirtzlin}, \citenamefont {Bocquet}, \citenamefont {Penkovsky}, \citenamefont
  {Klein}, \citenamefont {Nowak}, \citenamefont {Vianello}, \citenamefont
  {Portal},\ and\ \citenamefont {Querlioz}}]{hirtzlin2020digital}%
  \BibitemOpen
  \bibfield  {author} {\bibinfo {author} {\bibfnamefont {T.}~\bibnamefont
  {Hirtzlin}}, \bibinfo {author} {\bibfnamefont {M.}~\bibnamefont {Bocquet}},
  \bibinfo {author} {\bibfnamefont {B.}~\bibnamefont {Penkovsky}}, \bibinfo
  {author} {\bibfnamefont {J.-O.}\ \bibnamefont {Klein}}, \bibinfo {author}
  {\bibfnamefont {E.}~\bibnamefont {Nowak}}, \bibinfo {author} {\bibfnamefont
  {E.}~\bibnamefont {Vianello}}, \bibinfo {author} {\bibfnamefont {J.-M.}\
  \bibnamefont {Portal}}, \ and\ \bibinfo {author} {\bibfnamefont
  {D.}~\bibnamefont {Querlioz}},\ }\bibfield  {title} {\enquote {\bibinfo
  {title} {Digital biologically plausible implementation of binarized neural
  networks with differential hafnium oxide resistive memory arrays},}\
  }\href@noop {} {\bibfield  {journal} {\bibinfo  {journal} {Frontiers in
  neuroscience}\ }\textbf {\bibinfo {volume} {13}},\ \bibinfo {pages} {1383}
  (\bibinfo {year} {2020})}\BibitemShut {NoStop}%
\bibitem [{\citenamefont {LeCun}(1998)}]{lecun1998mnist}%
  \BibitemOpen
  \bibfield  {author} {\bibinfo {author} {\bibfnamefont {Y.}~\bibnamefont
  {LeCun}},\ }\bibfield  {title} {\enquote {\bibinfo {title} {The mnist
  database of handwritten digits},}\ }\href@noop {} {\bibfield  {journal}
  {\bibinfo  {journal} {http://yann. lecun. com/exdb/mnist/}\ } (\bibinfo
  {year} {1998})}\BibitemShut {NoStop}%
\end{thebibliography}
%


\end{document}